\documentclass{article}
\PassOptionsToPackage{numbers, compress}{natbib}
\usepackage[preprint] {neurips_data_2024}

\usepackage[utf8]{inputenc} % allow utf-8 input
\usepackage[T1]{fontenc}    % use 8-bit T1 fonts
\usepackage{hyperref}       % hyperlinks
\usepackage{url}            % simple URL typesetting
\usepackage{booktabs}       % professional-quality tables
\usepackage{amsfonts}       % blackboard math symbols
\usepackage{nicefrac}       % compact symbols for 1/2, etc.
\usepackage{microtype}      % microtypography
\usepackage{xcolor}         % colors
\usepackage{amsmath}
\usepackage{graphicx}
\usepackage{soul}
\usepackage{tabularx}
\usepackage{array}
\usepackage{caption}
\usepackage{xspace}
\usepackage{geometry}
\usepackage{longtable}
\usepackage{xcolor}
\usepackage{multirow}
\usepackage{colortbl}
\usepackage{subcaption}
\usepackage{algorithm}
\usepackage{algpseudocode}
\usepackage{pgfplots}
\usepackage{pgfplotstable}  % Make sure this is loaded
\pgfplotsset{compat=1.16}
\usepackage{enumitem}

\newcommand{\dbpedia}{\textsf{DBpedia}\xspace}
\newcommand{\lmdb}{\textsf{LMDB}\xspace}
\newcommand{\esbm}{ESBM\xspace}
\newcommand{\faces}{\textsf{FACES}\xspace}
\newcommand{\info}{\textsf{INFO}\xspace}

\newcommand{\am}[1]{\textcolor{black}{#1}}

\newcommand{\dm}[1]{{\color{black}{#1}}}

\newcommand{\ourdataset}{\textsc{WikES}\xspace}

\newcommand{\firstdataset}{\textsf{WikiLitArt}\xspace}
\newcommand{\seconddataset}{\textsf{WikiCinema}\xspace}
\newcommand{\thirddataset}{\textsf{WikiPro}\xspace}
\newcommand{\fourthdataset}{\textsf{WikiProFem}\xspace}

\newcommand{\comm}[1]{}

\newcommand{\xmat}{\mathrm{X}}
\newcommand{\smat}{\mathrm{S}}
\newcommand{\rmat}{\mathrm{R}}
\newcommand{\pmat}{\mathrm{P}}
\newcommand{\fmat}{\mathrm{F}}

\newcommand*{\KG}{\mathcal{KG}}
\newcommand{\R}{\mathcal{R}}

\newcommand{\V}{\mathcal{V}}
\newcommand{\E}{\mathcal{E}}

\newcommand{\T}{\mathcal{T}}
\newcommand{\K}{k}

\newcommand{\N}{\mathcal{N}}

\newcommand{\triple}[3]{(#1, #2, #3)}
\newcommand{\Nt}{\N_{t}}

\newcommand{\Desc}{\Delta}
\newcommand{\Su}{\mathcal{S}}
\DeclareMathOperator*{\argmax}{\arg\:\max}

\title{Wiki Entity Summarization Benchmark}

\author{%
  Saeedeh Javadi\thanks{Equal contribution} \\
  Polytechnic University of Turin \\
  \texttt{saeedeh.javadi@studenti.polito.it} \\
  \and
  \textbf{Atefeh Moradan}\textsuperscript{*} \\
  Aarhus University \\
  \texttt{atefeh.moradan@cs.au.dk} \\
  \and
  \textbf{Mohammad Sorkhpar}\textsuperscript{*} \\
  Indiana State University \\
   \texttt{msorkhpar@sycamores.indstate.edu}
  \and
  Klim Zaporojets \\
  Aarhus University \\
  \texttt{klim@cs.au.dk} \\
  \and
  Davide Mottin \\
  Aarhus University \\
  \texttt{davide@cs.au.dk} \\
  \and
  Ira Assent \\
  Aarhus University \\
  \texttt{ira@cs.au.dk} \\
}

\begin{document}

% Please read the instructions below carefully and follow them faithfully. \textbf{Important:} This year the checklist will be submitted separately from the main paper in OpenReview, please review it well ahead of the submission deadline: \url{https://neurips.cc/public/guides/PaperChecklist}.

\maketitle

\begin{abstract}

\dm{Entity summarization aims to compute concise summaries for entities in knowledge graphs. 
%Entity summarization is an important primitive for machine learning models and offers convenient solutions for critical applications, such as medical and biological domains. 
Existing datasets and benchmarks are often limited to a few hundred entities and discard graph structure in source knowledge graphs. This limitation is particularly pronounced when it comes to ground-truth summaries, where there exist only a few labeled summaries for evaluation and training. 
 We propose \ourdataset (Wiki Entity Summarization Benchmark), a comprehensive \emph{benchmark} comprising of entities, their summaries, and their connections. Additionally, \ourdataset features a dataset \emph{generator} to test entity summarization algorithms in different areas of the knowledge graph. 
 Importantly, our approach combines graph algorithms and NLP models, as well as different data sources such that \ourdataset does not require human annotation, rendering the approach cost-effective and generalizable to multiple domains. Finally, \ourdataset is scalable and capable of capturing the complexities of knowledge graphs in terms of topology and semantics. %This benchmark is instrumental in advancing research in semantic data and knowledge graphs, particularly in managing large datasets. 
\ourdataset features existing \emph{datasets} for comparison. Empirical studies of entity summarization methods confirm the usefulness of our benchmark. Data, code, and models are available at: \url{https://github.com/msorkhpar/wiki-entity-summarization}.}

%Entity summarization has recently gained importance, focusing on creating concise summaries for entities. Existing datasets often suffer from small size and bias. \ourdataset is a novel, fully automated benchmark and dataset generator for entity summarization using Wikidata and Wikipedia. This graph-based approach serves the NLP and graph theory by generating scalable datasets based on real-world data while preserving the topology of the Wikidata knowledge graph. Key features include automatic summary generation and the creation of datasets that reflect the structure of Wikidata. Baseline models were evaluated on the \ourdataset dataset in comparison to the ESBM dataset, demonstrating their efficiency and effectiveness. High quality annotations showed performance equivalent to human agreement, highlighting challenges such as scalability issues of current models. This benchmark offers significant potential for future research in semantic data processing and knowledge graph summarization. Data, code, and models are available at  \url{https://github.com/msorkhpar/wiki-entity-summarization} 

\end{abstract}

\comm{Entity summarization tasks has gained relevance in recent years. This tasks consists in .... Yet the existing datasets are too small and present biases. In this paper we introduce a novel entity summarization dataset. We show that this method is of high quality, obtaining equivalent to human performance annotation agreement. We expect that such large-scale dataset will present the community with new challenges. For example, we showcase that current models are unable to scale. As a solution we propose a model based on ... to tackle this. The relatively low performance of this state-of-the-art model opens the opportunity to research community to create more robust solutions to tackle entity summarization task. }
\section{Introduction}
% \begin{enumerate}
%     \item Paragraph 1: Contribution: scalable automatic dataset generation of entity summaries. Highly challenging real-world dataset, with (we expect) little bias. The entities are chosen from multiple domains randomly (given that they contain wikipedia pages with abstracts). To do this, we embed the text, and chose the entities that are highly dispersed (not clustered). This way, ensuring that the entities are not biased towards a particular topic. Use of semantic similarity. 
%     \item Paragraph 2: present statistics and main challenges. (We expect) low frequency bias, due to highly heterogeneous domains. The naive models can perform very good. 
%     \item Paragraph 3: the high quality of dataset/annotator agreement. We use 3 expert annotators, calculate the inter-annotator agreement and then the agreement with the produced annotations. 
% \end{enumerate}

\dm{\textit{Knowledge Graphs} (KGs) are a valuable information representation: interconnected networks of entities and their relationships enable machine reasoning to empower question answering \cite{hu2018answering, lan2019multihop}, recommender systems \cite{wang2018dkn}, information retrieval \cite{raviv2016document}. KGs may comprise millions of entities representing real-world objects, concepts, or events.}

\begin{figure}[h]
    \centering
    \includegraphics[width=10cm]{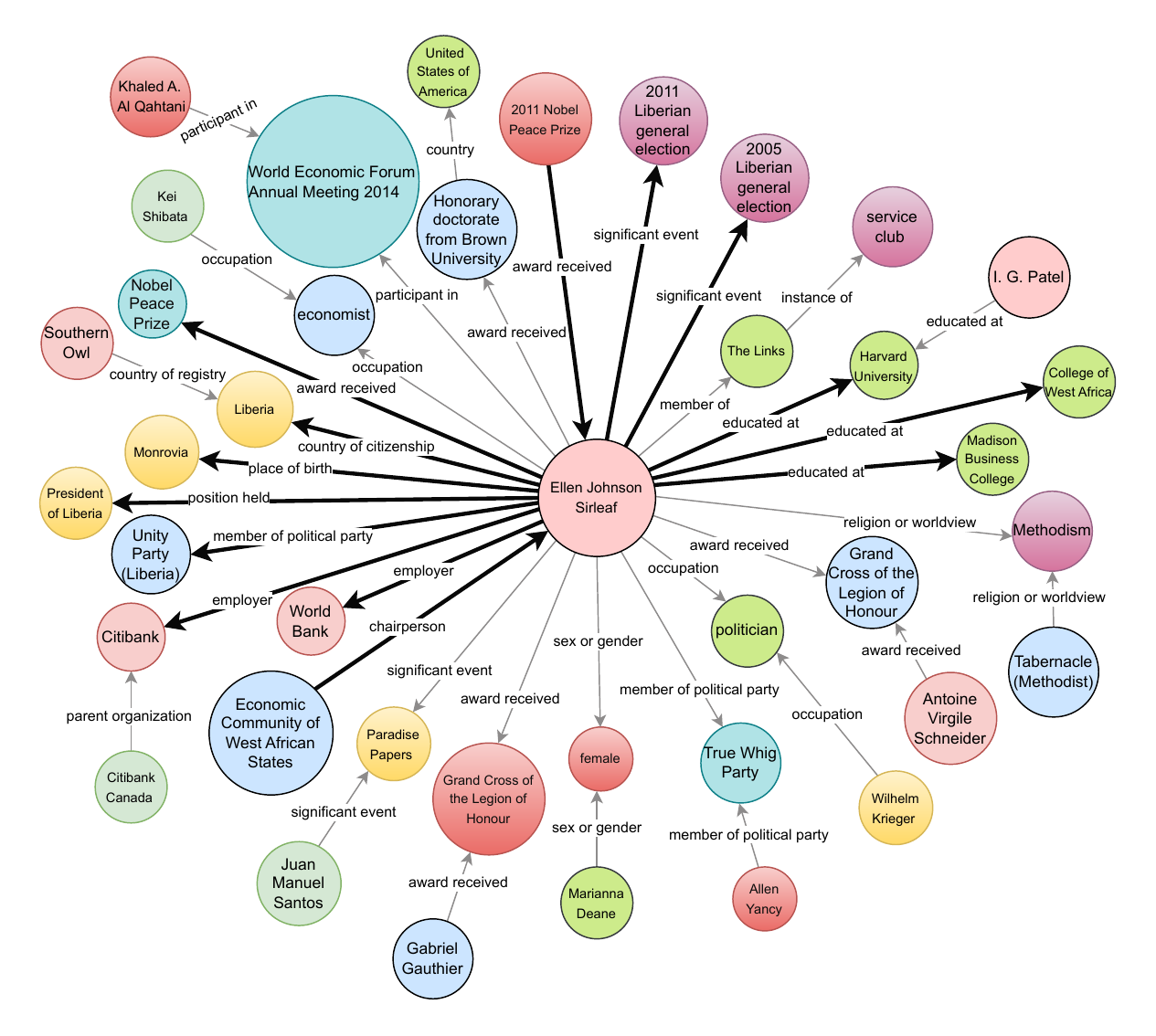}
    \caption{KG subgraph of entity \texttt{Ellen Johnson Sirleaf}: arrows depict the subgraph of relationships to other entities, and labels indicate their roles. Selecting the bold edges as entity summaries of the most relevant triples may reduce information overload while concisely describing the entity. }
    \label{fig:EllenJohnsonSirleaf}
\end{figure}

\dm{Yet, the size and complexity of these KGs progressively expand, rendering it increasingly challenging to convey the essential information about an entity in a concise and meaningful way~\cite{suchanek2007yago, vrandevcic2014wikidata}. This is where entity summarization becomes relevant. \textit{Entity summarization} (ES) \cite{liu2021entity} is the process of generating a concise and informative summary that captures the most salient aspects of the entity description, based on the information available in the KGs. In ES, the entity \emph{description} refers to all the triples involving such an entity. For instance, Figure~\ref{fig:EllenJohnsonSirleaf} illustrates a set of relationships surrounding the entity \texttt{Ellen Johnson Sirleaf} in a KG, along with a possible summary for this entity. Extensive descriptions can overwhelm users and exceed the capacity of typical user interfaces, making it challenging to identify the most relevant triples. Entity summarization addresses this issue by computing an optimal compact summary for an entity, selecting a size-constrained subset of triples \cite{liu2021entity}. }

\dm{Despite advances in entity summarization techniques~\cite{liu2021entity}, the development and evaluation of these methods are hindered by a number of limitations in the benchmarks and datasets~\cite{liu2020esbm,cheng2023characteristic}.
The first limitation of the current benchmarks is the small dataset size, encompassing only a few hundred entities. Second, the generation of ground-truth summaries for testing mostly relies on expensive and lengthy manual annotation. Moreover, the dependence on a few human annotators often biases the data towards the annotators' preferences and knowledge. Third, existing benchmarks often disregard the wealth of information in the knowledge graph structure.}

To address the above limitations, we propose: 
\begin{itemize}[leftmargin=*, noitemsep]
    \item\dm{\textbf{Novel \ourdataset benchmark for ES} based on summaries and graphs from Wikidata and Wikipedia.}
    %yCreating datasets that closely mirror the topology of the Wikidata graph, accurately replicating its overall structure.
    \item \dm{ \textbf{Subgraph extraction method} preserving the complexity of real-world KGs; subsampling using random walks and proportionally preserving node degrees, \ourdataset captures the structure of the entities up to the second-hop neighborhood, thereby ensuring that the connections in \ourdataset accurately reflect those in the source KG. }
    \item \dm{ \textbf{Comprehensive summaries for \emph{any} entity in the KG}, ensuring that summaries are both relevant and contextually rich by deriving them directly from corresponding Wikipedia abstracts,  minimizing human bias, as these abstracts are created and reviewed by several experts. In this manner, \ourdataset is scalable, enabling it to generate large benchmark resources efficiently with high-quality annotation.}
    \item \dm{\textbf{Automatic entity summarization dataset generator} allows for the creation of arbitrarily large datasets, encompassing various domains of knowledge. }
\end{itemize}

\comm{\revklim{Story: 
Contributions: 
\begin{itemize}
    \item \ul{Scalable, automatic, high quality and free} dataset generation method for entity summarization. Generated datasets for number of entities: 
    \begin{itemize}
        \item \ul{Scalable}: datasets generated for small (1,000), medium (10,000) and large (100,000) root entities. 
        \item \ul{Automatic}: easily executable interface, automatic sampling of root entities (cross-domain, based on click/wikipedia page length), or passing root entities as a list. 
        \item \ul{High quality/reliable way of annotating}: annotator agreement on a subsample of 100 entities. 
        \item \ul{Free (open)}: open license, based on Wikipedia/Wikidata. 
    \end{itemize}
    \item Evaluation of baseline methods: 
    \begin{itemize}
        \item Traditional statistic-based methods (unsupervised).
        \item Supervised (GNNs?) based methods. 
        \item Others? 
    \end{itemize}
    \item Statistics comparing to related datasets. 
    \item Analysis of metrics for not-fixed number of candidates; here we also differ from ESBM. 
    \begin{itemize}
        \item \textbf{Rank-precision metric (RP)}
        \item \textbf{Average Precision}
    \end{itemize}
\end{itemize}
}}

\section{Existing Datasets}

\am{Here, we review the existing datasets for entity summarization. Table~\ref{tbl:existing-datasets} provides an overview and statistics of the current datasets in this field. \faces and \info datasets have a higher density than the entities in the Entity Summarization Benchmark (\esbm). It is also clear that \lmdb and \faces are not connected graphs, that challenge graph-based learning methods where the information cannot easily propagate in disconnected networks. Specifically, \faces consists of 12 connected components, which complicates the learning process for graph embedding methods by limiting the richness of information that can be leveraged from the graph.}

\begin{table}[!ht]
\footnotesize % Reduce font size
\caption{\dm{Entity summarization datasets in terms of number of entities $|\V|$, triples $|\E|$, number of ground-truth summaries (target entities), density as $\displaystyle|\E|/\binom{|\V|}{2}$, graph connectivity, number of components, sampling method to select entities and subgraph, and minimum / maximum node degree.}}
\begin{tabularx}{\linewidth}{Xrrrr} % Adjusted column definition
\toprule
Metric & \dbpedia (\esbm)  & \lmdb (\esbm) & \faces & \info \\ 
\midrule
Entities ($|\V|$) & 2\,721 & 1\,853 & 1\,379  & 1\,410  \\ 
Relations ($|\E|$) & 4\,436 & 2\,148 & 2\,152  & 2\,019   \\ 
Target Entities & 125 & 50 & 50 & 100  \\ 
Density & 0.0005 & 0.0006 &  0.0011  &  0.0010 \\ 
%Density (Undirected) & 0.0011 & 0.0011 & 0.0022  & 0.0020 \\ 
Sampling method & Not specified & Not specified &  Not specified  & Not specified  \\ 
Connected-graph & Yes & No & No & Yes  \\ 
Num-comp & 1 & 2 &  12  &  1 \\ 
Min Degree & 1 & 1 &  1  &  1 \\ 
Max Degree & 125 & 208 &  88  &  100 \\ 
\bottomrule
\end{tabularx}
\label{tbl:existing-datasets}
\end{table}

% The only dataset with comparison table I have seen so far is: MBW: Multi-view Bootstrapping in the Wild
%\input{sections/introduction} 

\dm{We provide here a comprehensive description of each dataset or benchmark:}

\begin{itemize}[leftmargin=*,noitemsep]
    \item\dm{\textbf{\esbm}~\cite{liu2020esbm}: The Entity Summarization Benchmark (\esbm) is the first benchmark to evaluate the performance of entity summarization methods. ESBM has three versions; v1.2 is the latest and most extensive version. This version comprises 175 entities, with 150 from DBpedia \cite{lehmann2015dbpedia} and 25 from LinkedMDB \cite{hassanzadeh2009linked}. The summaries comprise triples selected by $30$ ``researchers and students`` annotators. Each entity has exactly $6$ summaries.  Despite encompassing two datasets, ESBM has several limitations. First, the entity sampling method is not explained. In particular, some triples in the neighborhood of the entity are missing in the datasets. Second, there are no connections among the entities in the neighborhood,  nor any two-hop neighborhood. Third, the expertise and background of the annotators are not assessed nor disclosed. 
    Due to the expensive annotation process, the dataset size is small.}
    \item \dm{ \textbf{FACES}~\cite{gunaratna2015faces}
    is a dataset from DBpedia (version 3.9)~\cite{auer2007dbpedia} and includes $50$ randomly selected entities, each with at least $17$ different types of relations. Similar to ESBM, the FACES ground-truth is also generated manually.}
    %The FACES dataset is used to evaluate the diversity-aware entity summarization approach. It is derived from DBpedia (version 3.9) and includes 50 randomly selected entities, each having at least 17 distinct properties. The average number of distinct features per entity is 44. The dataset focuses on object-type properties to provide cleaner and more informative data for manual evaluation. This setup ensures that the dataset captures a wide range of information about the entities, though the dataset's reliance on object-type properties and the manual creation of gold-standard summaries introduce potential biases and limit scalability.
    %is a subset of the English version of DBpedia~\cite{auer2007dbpedia} that uses seven core datasets from DBpedia, namely the Ontology Infobox Types, Ontology Infobox Properties, Titles, Geographic Coordinates, Homepages, Persondata, and PND. The datase
    \dm{\item \textbf{INFO}~\cite{cheng2023characteristic} contains $100$ randomly selected entities from $10$ classes in DBpedia. It comprises two sets of ground-truth summaries, REF-E and REF-W. REF-E summaries comprise a selection of triples from five experts adhering to a 140-character limit, similar to typical Google search result snippets. REF-W summaries are obtained by one expert who reads the abstract sections of the respective entities on Wikipedia and selects neighboring entities that closely match the Wikipedia abstracts. 
    The number of ground-truth summaries per entity varies, as some experts evaluate multiple entities. This inconsistency complicates the evaluation process. The expertise of the annotators remains unspecified.}
\end{itemize}

\am{In contrast, our benchmark uses Wikidata to automatically map entities from Wikipedia to Wikidata. This automation allows us to efficiently generate summaries for any number of entities. Unlike previous work, we use the Wikipedia abstract as a summary instead of manual annotators. Each abstract is a collaboration of many users; as such, it should not introduce obvious biases. Additionally, with this process, we ensure high-quality and cost-effective summaries. Furthermore, we present the characteristics of our dataset in Table~\ref{tbl:combined-dataset}. The \ourdataset benchmark includes a larger number of entities and relations than existing datasets. It is a connected graph containing approximately 500 seed nodes. Further details regarding the specific characteristics of our dataset are provided in Section~\ref{sec:dataset-info}.}

%extracts structured encyclopedic information from Wikipedia using an RDF graph  making it a popular dataset for knowledge graph completion (KGC). 3.7 that contains several million entities and a total of 42.3 million RDF triples.  The dataset's diversity and the structured nature of the data make it suitable for evaluating entity summarization methods. transforming high-quality textual summaries into sets of features.  This dual approach provided a robust benchmark for assessing the performance of summarization methods, with the summaries showing high agreement among experts.

\section{The \ourdataset Benchmark}
\label{sec:data-collect}

\dm{A \emph{Knowledge Graph} $\KG = (\V, \R, \T)$ is a directed multigraph consisting of entities $\V = \{v_1, \dots, v_n\}$, relationships $\R$, and triples $\T \subseteq \V \times \R \times \V$. The set of edges $\E = \{(i, j) \mid v_i, v_j \in \V \wedge \exists r \in \R \text{ s.t. } \triple{v_i}{r}{v_j} \in \T\}$ contains pairs of nodes connected by a relationship.}
% We extend the entity summarisation problem~\cite{liu2021entity} to incorporate nodes and edges beyond an entity's direct neighbours, harnessing the valuable descriptive information they provide. We first define the direct neighbourhood and then generalize it to include neighbours connected through multiple edges.

The \emph{$t$-hop neighborhood} $\Nt(v_i)$ of node $v_i$ is the set of nodes reachable from $v_i$ within $t$ edges when ignoring edge directions. 
% Note that our definition is a straightforward generalisation of \cite{liu2021entity} in which $\Desc$ contains only the immediate neighbors; thus $\Desc$ is identical to $\Desc_1$. 

% The problem of entity summarisation is then one of selecting information from the available neighborhood information: 

A \emph{summary} for an entity $v_i$ is a subset $\Su(v_i) \subseteq \Desc_{t}(v_i)$ of triples from the $t$-description of $v_i$, where the $t$-\emph{description} of an entity $v_i \in \V$ in a knowledge graph $\KG$ is the set $\Desc_t(v_i) = \{ \triple{s}{p}{o} \in \T \mid s \in \Nt(v_i) \vee o \in \Nt(v_i) \}$ of triples in which one of the entities is in the $t$-hop neighborhood of $v_i$.

% In entity summarisation, we wish to find a summary that extracts the most important information from the description, as measured by an importance score:

\textbf{Entity summarization} for an entity $v_i\in \V$ in a knowledge graph $\KG$ aims to find a summary $\Su(v_i)$ that maximizes some score among all possible summaries for $v_i$, i.e., 
\begin{equation}
\underset{\substack{\Su(v_i) \subseteq \Desc_{t}(v_i)\\|\Su(v_i)|=\K}}\argmax {\mathrm{score}(\Su(v_i))}, 
\end{equation}

% \am{The \emph{$t$-hop neighbourhood} $\Nt(v_i)$ of node $v_i$ is the set of nodes reachable from $v_i$ within $t$ edges, ignoring edge directions. 
% The $t$-\emph{description} of an entity $v_i \in \V$ is $\Desc_t(v_i) = \{ \triple{s}{p}{o} \in \T \mid s \in \Nt(v_i) \vee o \in \Nt(v_i) \}$. A \emph{summary} for an entity $v_i$ is a subset $\Su(v_i) \subseteq \Desc_{t}(v_i)$. \emph{Entity summarization} for $v_i\in \V$ aims to find a summary $\Su(v_i)$ that maximizes a score:
% \begin{equation}\label{eq:entity_sum}
% \underset{\substack{\Su(v_i) \subseteq \Desc_{t}(v_i)\\|\Su(v_i)|=\K}}\argmax {\mathrm{score}(\Su(v_i))}.
% \end{equation}}

%\subsection{ \ourdataset Generator Algorithm}

%\am{The \ourdataset process involves two steps: first, preprocessing, and second, dataset generation.}

%The first step in preprocessing is loading Wikipedia and Wikidata into the databases for further processing. In this phase,

\subsection{Extracting Summaries from Wikidata using Wikipedia Abstracts}
\label{sec:preprocess}
\dm{We extract summaries for each Wikidata item using Wikipedia abstracts and infoboxes. Each abstract is a joint effort of many users and experts, which ensures quality and accuracy. Leveraging Wikipedia, we avoid time-consuming manual annotation and enable the automatic generation of large-scale datasets.} 

%\am{We use Java version 21, Spring Boot version 3, Docker version 24.0.8, Python version 3.10, PostgreSQL version 16.3, Neo4j version 5.20.0-community, Wikpedia XML article dump files published on 2023/05/01, and the Wikidata XML article dump files published by Wikimedia on 2023/05/01.}

%For the pre-processing steps, we used an r5a.4xlarge instance on AWS with the following specifications: vCpu: 16 (AMD EPYC 7571, 16 MiB cache, 2.5 GHz) Memory: 128 GB (DDR4, 2667 MT/s)Storage: 500 GB (EBS, 2880 Max Bandwidth)

%As referenced in Section~\ref{sec:data-collect}, Wikidata consists of items , statements (triples), and properties (i. Wikidata items are the subjects of the triples, which correspond to other entities (nodes) $\V$ in the $\KG$ definition. Wikidata triples are the relations (edges) $\R$ in the $\KG$. Wikidata properties are the labels over the connecting edges that connect Wikidata items. Furthermore, each Wikidata item and property contains a label and a description.

%To be able to process these sections we extract the latest reversion of a page and save it into our database.

\dm{\textbf{Wikidata} is a free and collaborative knowledge base that collects structured data to support Wikipedia and other Wikimedia projects. It includes descriptions and labels for entities. The descriptions offer in-depth details, while the labels serve as concise identifiers, facilitating efficient data retrieval and integration in subsequent steps. 
We load all Wikidata items XML dump files published on 2023/05/01\footnote{\url{https://dumps.wikimedia.org/wikidatawiki/}} as entities $\V$ alongside their properties as relationships $\R$ into a graph database\footnote{\url{https://neo4j.com}}. The result is a graph that connects all Wikidata items and statements. We include items if they (1) are not marked as redirects, (2) belong to the main Wikidata namespace, and (3) have an English label or description. Additionally, we load metadata for each Wikidata item and property, including labels and descriptions, into a relational database\footnote{\url{https://www.postgresql.org/}}.}
%If a Wikidata item has a link to a Wikipedia page, we  store such a link.}
% \dm{\textbf{Wikidata} consists of Wikidata items (identifiers, starting with the letter 'Q') and Wikidata property identifiers, starting with the letter 'P'). 

\dm{\textbf{Wikipedia} pages contain infoboxes, abstracts, page content, categories, references, and more. Links to other Wikipedia pages are referred to as mentions. We detect these mentions in the abstracts and infoboxes of Wikipedia pages to use them later for labeling the summaries in Wikidata.}
\am{We extract and load all the content from the XML dump files of Wikipedia pages, published on 2023/05/01\footnote{\url{https://dumps.wikimedia.org/enwiki/}}, into a relational database under the same conditions as Wikidata: the pages must be in English and not redirected.}

\dm{\textbf{Summary annotation.}
We annotate the summaries in Wikidata using the corresponding Wikipedia pages. 
For each Wikipedia page corresponding to a Wikidata entity, we iterate through all connected Wikidata items using Wikidata properties. If a connected Wikidata item is mentioned in the Wikipedia abstract and infobox, we annotate the Wikidata item with the corresponding Wikidata property as part of the summary.}

\dm{Wikidata is a directed multigraph, which means that each entity (Wikidata item) can be connected to another entity via multiple relations (Wikidata properties). Yet, links in Wikipedia are not labeled; as such, we need to select one of the relations for the summary. To annotate the correct Wikidata property as part of the summary, we employ the DistilBERT model~\cite{Sanh2019DistilBERTAD}. DistilBERT is a fast and lightweight model with a reduced number of parameters compared to the original BERT model. This way, we can efficiently process large amounts of data while maintaining high-quality embeddings for accurate relation selection.}

\dm{Concretely, we first embed the abstract of the Wikidata item for which we are generating summaries using DistilBERT. We then calculate the cosine similarity between the embedding of the abstract and the embeddings of each candidate relation. Finally, we add the relation with the highest cosine similarity to the abstract embedding to the summary.}
\dm{This approach ensures that the most relevant Wikidata property is selected for the summary based on its semantic similarity to the Wikipedia abstract.}

%\mks{ We iterate the pages and for the pages mentioned in the current page abstract and infobox, we annotate the relation as a summary if there exists a relation connecting the current page and the mentioned page together in Wikidata. As we know Wikidata is a multigraph that edges in the $\KG$ might have multiple labels connecting a pair of nodes in it. To pick the best predicate(label) out of the candidates we found while searching for the edges, we use the DistiBert NLP model. We first use the abstract for embedding and then other embedding over the predicates found as candidates. Having these two embeddings, we run a cosine similarity of each candidate and the abstract and pick argmax of the result as the most relevant candidate for connection between the page and mention. When all the pages are annotated, we are done with the preprocessing, and by inquiring about a set of seed nodes, we can get a graph containing all the neighbors that are considered as summaries for that target node.   

%\textbf{Mapping Summaries of Wikidata and Wikipedia}

\subsection{Capturing the Graph Structure}
\label{algo:randwalk}

\dm{Here we introduce the \ourdataset generator algorithm. The main idea is to sample a connected graph that preserves the original graph structure. To this end, we employ random walks~\cite{pearson1905randomwalk}. }

\dm{A random walk is a stochastic process defined as a sequence of steps, where the direction and magnitude of each step are determined by the random variable 
$\xmat_{t+1} = \xmat_t + \smat_t$
where \(\xmat_t\) represents the position at time \(t\), and \(\smat_t\) is the step taken from position \(\xmat_t\).}

\dm{The process is a Markov process, characterized by its memoryless property:
\begin{equation}
P(X_{t+1} = x | X_t = x_t, X_{t-1} = x_{t-1}, \dots, X_0 = x_0) = P(X_{t+1} = x | X_t = x_t)
\end{equation}}
\am{In adapting this concept to our work, we redefine the number of random walks assigned to nodes based on their degrees, ensuring the distribution remains proportional to real data. This is achieved through logarithmic transformation and normalization. The logarithmic transformation is applied to reduce the impact of high-degree nodes and also low-degree nodes, making it more manageable for the random walk. Given a graph with node degrees \( \{d_1, d_2, \ldots, d_i\} \),  the log-transformed degree for node $i$ is $L_i = \log(d_i)$. These values are then normalized:
\begin{equation}
N_i = \frac{L_i - \min(\{L\})}{\max(\{L\}) - \min(\{L\})}
\end{equation}
where \(N_i\) is the normalized logarithmic degree of node \(i\). Finally, the number of random walks \(R_i\) assigned to each node is: 
\begin{equation}
R_i = \text{round}\left(\text{minRW} + N_i \times (\text{maxRW} - \text{minRW})\right)
\end{equation}}
\dm{Here, \(\text{minRW}\) and \(\text{maxRW}\) are the user-defined minimum and maximum limits for random walks. This adaptation ensures that the random walks are proportional to the normalized logarithmic degree of each node, reflecting the true structure of the network.
For a small dataset we set \(\text{minRW} = 100\) and \(\text{maxRW} = 300\); for a medium dataset \(\text{minRW} = 150\) and \(\text{maxRW} = 600\); for a large dataset, \(\text{minRW} = 300\) and \(\text{maxRW} = 1800\). This ensures that the random walks are tailored to both the scale and the complexity of the dataset.} \am{Importantly, our approach can be used to extract further subgraphs at the scale needed for benchmarking in a given scenario.}

\am{Moreover, the random walk sampling process requires a set of seed nodes as a starting point. In our case, the seed nodes represent the target entities we are interested in. The seed nodes can be any Wikidata Item Identifier, Wikipedia title, or Wikipedia ID of the Wikipedia pages. We collect the seed nodes on the condition that they have at least $k$ (default $k=5$) common entities with the abstract section and the infobox in the Wikipedia pages. Therefore, this model is flexible, allowing you to choose any seed nodes from any domain as an input.
%\am{Collecting quality seed nodes is challenging since they should be related and have well-written abstracts and infoboxes, which can limit the scope of the collected data.}
In the datasets that we generated, we collect seed nodes from~\cite{laouenan2022cross}. This paper has published information about individuals from various domains. The authors collected data from multiple Wikipedia editions and Wikidata, using deduplication and cross-verification techniques to compile a database of 1.6 million individuals with English Wikipedia pages. The seed nodes that we use include actor, athletic, football, journalist, painter, player, politician, singer, sport, writer, lawyer, film, composer, novelist, poet, and screenwriter.}
\dm{Using combinations of these seed nodes, we generate four sets of datasets, with each set having small, medium, and large versions. In Table~\ref{tab:seed_nodes} in Section~\ref{sec:appendix} in the supplementary material, we present the seed nodes and their proportions for each dataset and their corresponding train-test-val splits.}
\label{sec:seed-node}
\subsection{\ourdataset Generator}
\am{We discuss how \ourdataset is created, and how further benchmarks can be generated without the need for manual annotators.} 
\dm{Algorithm~\ref{alg:generator} details the generator, which consists of the following steps.}

\dm{\textbf{Step1:} Retrieve summaries of each seed node (explained in Section~\ref{sec:preprocess})}

\dm{\textbf{Step2:} Expand the graph using the random walk method in Section~\ref{algo:randwalk}. Set the random walk's length $n$ (default $n=2$), which means it explores up to the $n$-hop neighborhood of each seed node.}

\dm{\textbf{Step3: }  Check if the graph is connected. If it is, done. If not, identify all disconnected components and sort them by size, from largest to smallest. In each iteration, connect smaller components to the largest component using $h$ connections. Utilize the shortest path method, selecting paths that are equal to or less than a minimum path length $l$. Continue connecting nodes from the smaller component to the larger one until $h$ nodes are connected. After each iteration, check graph connectivity again. If all components are connected to the largest component, the algorithm ends. Otherwise, re-sort components and increase $l$ by 1. Repeat until the graph is a single connected component.}
\begin{algorithm}
\caption{\ourdataset Generator}
\label{alg:generator}
\begin{algorithmic}[1] 
\footnotesize
\State \textbf{Input:} Graph $G$, seed nodes $S$, random walk length $n$, minimum path length $l$
\State \textbf{Output:} A connected graph
\Procedure{GenerateGraph}{$G$, $S$, $n$, $l$}
    \State $\textit{summaries} \gets \Call{RetrieveSummaries}{S}$
    \State $G \gets \Call{RandomWalkExpansion}{G, S, n}$ mentioned in section~\ref{algo:randwalk}
    \State $is\_connected \gets \Call{CheckConnectivity}{G}$
    \While{not $is\_connected$}
        \State $components \gets \Call{FindComponents}{G}$
        \State Sort $components$ by size in descending order
        \State $largest \gets components[0]$
        \For{$comp$ in $components[1:]$}
            \State Connect $comp$ to $largest$ using $h$ connections via shortest paths $\leq l$
            \State $G \gets \Call{UpdateGraph}{G, comp, largest}$
            \State $is\_connected \gets \Call{CheckConnectivity}{G}$
            \If{$is\_connected$}
                \State \textbf{break}
            \EndIf
        \EndFor
        \State $l \gets l + 1$
    \EndWhile
    \State \textbf{return} $G$
\EndProcedure
\end{algorithmic}
\end{algorithm}

%\subsection{Seed Node Collection}
%\input{tables/seed-node-tbl}
% We construct four datasets of varying sizes using the seed nodes described in Section~\ref{sec:seed-node} and random walks.}
%\textbf{\ourdataset Meduim} We have generated four small datasets. We present their characteristics in Table~\ref{tbl:our-dataset-small}. The number of entities in the small datasets ranges from approximately 70$k$ to 85$k$, and the number of relations ranges from around 120$k$ to 135$k$
%\textbf{\ourdataset Large} We have generated four small datasets. We present their characteristics in Table~\ref{tbl:our-dataset-small}. The number of entities in the small datasets ranges from approximately 70$k$ to 85$k$, and the number of relations ranges from around 120$k$ to 135$k$
\subsection{\ourdataset Datasets}
\label{sec:dataset-info}
\am{We generate three sizes for each of the four datasets, obtaining 12 datasets. We present their characteristics in Table~\ref{tbl:combined-dataset} in section ~\ref{sec:appendix}. 
The number of entities in the small datasets ranges from approximately 70$k$ to 85$k$, and the number of relations ranges from around 120$k$ to 135$k$. In the medium datasets, the number of entities ranges from 100$k$ to 130$k$, and the number of relations ranges from 195$k$ to 220$k$. The number of entities in the large datasets ranges from approximately 185$k$ to 250$k$, and the number of relations ranges from around 397$k$ to 470$k$.
The average runtime for generating small graphs is approximately $128$ seconds; for medium-sized graphs, it is approximately $216$ seconds; and for large graphs, it is approximately $512$ seconds. 
We construct the train-test-validation split for each dataset with \(70\%\) for training, \(15\%\) for testing, and \(15\%\) for validation.  Detailed information about the run time, as well as the number of nodes and relations for these splits, is available on our GitHub repository. All graphs in each train-test-validation splits are connected.}
\section{Empirical Evaluation}
\label{sec:evaluation}
We study the quality of \ourdataset using the following metrics:

{\textbf{F-Score.}} \dm{Let $\Su_{m}$ the summary obtained by a summarization method and $\Su_{h}$ the ground-truth summary. We compare $\Su_{m}$ with $\Su_{h}$ using the F1-score based on precision $P$ and recall $R$:
\begin{equation} \label{eq:f1}
    \begin{aligned}
    \fmat1 = \frac{2{\cdot}P{\cdot}R}{P+R},\ 
    \text{where}\ 
    \pmat = \frac{|\Su_{m} \cap \Su_{h}|}{|\Su_{m}|}\ 
    \text{and}\ 
    \rmat = \frac{|\Su_{m} \cap \Su_{h}|}{|\Su_{h}|}
    \end{aligned}
\end{equation}
The F1 score lies within [0,1]. High F1 indicates that $\Su_{m}$ is closer to the ground-truth $\Su_{h}$.}

{\textbf{Mean Average Precision (MAP).}} \dm{This metric is particularly suitable for evaluating ranking tasks because it takes into account the order of the predicted triples. MAP calculates precision at each position $i$ in the predicted summary and averages these values over all relevant summary triples. It reflects both the relevance and the ranking quality of the predicted summaries. MAP, unlike F1-score, does not depend on a specific value of $k$. This makes it a robust metric for assessing how well a summarization method ranks the relevant triples.
\begin{equation}
    \mathrm{MAP} = \frac{1}{N} \sum^{N}_{n=1} \dfrac{\sum^{|\Su_{m}^{(n)}|}_{i=1} \begin{cases}
\mathrm{Precision}@i(\Su_{h}^{(n)}) & \mathrm{if}\ \mathrm{Rel}(n, i)\\
0 & \mathrm{otherwise}
\end{cases}}{|\Su_{h}^{(n)}|}
\end{equation}
where $N$ is the total number of entities, $\Su_{h}^{(n)}$ is the set of ground-truth summary triples for a particular entity $v_n$, $\Su_{m}^{(n)}$ is the set of predicted summary triples for the entity $v_n$, $\mathrm{Precision}@i$ is the precision at the $i$-th position in the predicted summary, and $\mathrm{Rel}(n, i)$ indicates whether the $i$-th predicted triple for entity $v_n$ is relevant (i.e., it belongs to $\Su_{h}^{(n)}$). MAP scores are in the range [0,1], where a higher MAP indicates better performance in terms of correctly predicting relevant summary triples.} \am{To account for the varying lengths of the ground-truth summaries in real-world data, we also calculate MAP and F-score (which we refer to as dynamic MAP and dynamic F-score) by setting the length of the generated summary ($|\Su_{m}|$) equal to the length of the corresponding ground-truth summary ($|\Su_{h}|$).}

%\subsection{Assessing Dataset Quality }
%\am{Here is the revised version of the text:
\dm{We analyze our dataset and compare it with the \esbm benchmark using statistical measures such as frequency and inverse frequency of entities and relations. We calculate the F-score and MAP score for the top-5 and top-10 of both the \esbm dataset and our \fourthdataset. We choose top-5 and top-10 because we only have ground-truth summaries for top-5 and top-10 in the \esbm dataset.}
\am{The F-score and MAP results for \esbm are presented in Figure~\ref{fig:esbm-stats}. The statistics show that for \dbpedia, the F-score using inverse relation frequency outperforms the random baseline by 0.15 for top-5 and by 0.34 for top-10. Furthermore, when using inverse entity frequency, \dbpedia achieves an even higher F-score, surpassing the random baseline by 0.07 for top-5 and by 0.15 for top-10.
%\dbpedia achieves a higher F-score than a random summary in terms of inverse entity frequency and inverse relation frequency
For \lmdb, we observe a similar trend when using inverse frequency. The F-score surpasses the random baseline by 0.10 for top-5 and by approximately 0.15 for top-10. Additionally, when employing entity frequency, \lmdb achieves an F-score that is around 0.17 higher than the baseline for top-5 and 0.07 higher for top-10.
The results demonstrate that \esbm exhibits a strong bias towards entity, reverse entity, and relation frequency. For Map score, we are exactly observing the same behavior for \esbm. We believe that the bias comes from the fact that the datasets are small, their second-hop neighborhood is not considered, and the relations between their first-hop neighbors are not considered.}
%On the other hand, \ourdataset in Figures~\ref{fig:f1-wikies} and~\ref{fig:map-wikies}. 
On the other hand, Figure~\ref{fig:f1-wikies} shows the F-score for top-$5$, top-$10$  and dynamic F-score on \ourdataset. Since the length of summaries varies with the abstract, we calculate the F-score for each seed node based on its summary length. Results show that \ourdataset F-score is close to random for different statistics, thus rejecting the hypothesis of obvious biases. We observe a minor bias towards node frequency in small datasets. Yet, as we increase the size of the dataset, this bias disappears. We observe a similar behavior with MAP in Figure~\ref{fig:map-wikies}
Furthermore, we use \emph{the entire} Wikidata to measure the F-score for our seed nodes. Thus, importantly, we observe that our dataset's F-score trend is comparable to that of the entire data, especially our large dataset. We also extracted the first-hop neighborhood of all our seed nodes and observed a small bias in the F-score top-5 and dynamic F-score. We conclude that adding the two-hop neighborhood makes the sample follow the graph distribution.
Thus, \ourdataset is an unbiased benchmark that retains the source KG distribution.

%\am{However, in \dbpedia, the F-score using inverse relation frequency for top-5 is 0.15 more than random and 0.34 more than random for top-10. In \lmdb, we also observe the same behavior for inverse frequency; it is 0.10 more than random in top-5 and around 0.15 more than random in top-10. The results show a strong bias, and we believe that the bias comes from the fact that the datasets are small and their second-hop neighborhood and the relations between their first-hop neighbors are not considered. For Map score we are excatly observing the same behaviour for \esbm datasets and \fourthdataset. }

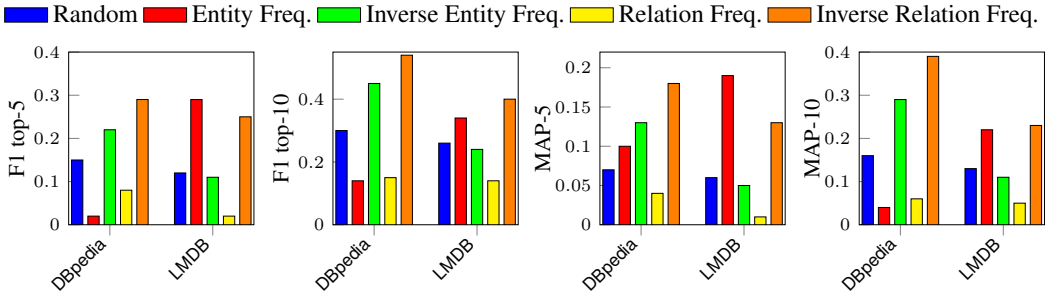
\begin{figure}[htb]
\centering
% First subfigure
\begin{subfigure}[b]{0.249\textwidth}  % Adjusted width
\centering
\begin{tikzpicture}[scale=0.95] 
\begin{axis}[
hide axis,
xmin=0,
xmax=10,
ymin=0,
ymax=10,
legend columns=5,
legend style={at={(0.5,1)},anchor=north,draw=none,fill=none,legend cell align=left, xshift=2cm},
legend image post style={scale=1},  % Add this line to scale down the legend images
legend entries={Random, Entity Freq., Inverse Entity Freq., Relation Freq., Inverse Relation Freq.}
]
\addlegendimage{fill=blue,area legend}
\addlegendimage{fill=red,area legend}
\addlegendimage{fill=green,area legend}
\addlegendimage{fill=yellow,area legend}
\addlegendimage{fill=orange,area legend}
\end{axis}
\end{tikzpicture}
\begin{tikzpicture}[scale=0.95] 
\begin{axis}[
ybar,
ylabel={\footnotesize F1 top-5},
ylabel style={font=\footnotesize, yshift=-0.5em},
ymin=0,
ymax=0.4,
width=1.2\textwidth,
bar width=1ex,
height=4cm,
xtick=data,
xticklabels={\dbpedia, \lmdb},
xtick pos=bottom,
ytick pos=left,
enlarge x limits=0.4,
yticklabel style={font=\scriptsize, /pgf/number format/fixed},
tick label style={font=\scriptsize},
xticklabel style={rotate=45, anchor=east}
]
\addplot [fill=blue]   coordinates {(0,0.15) (1,0.12) };
\addplot [fill=red]    coordinates {(0,0.02) (1,0.29)};
\addplot [fill=green]  coordinates {(0,0.22) (1,0.11) };
\addplot [fill=yellow] coordinates {(0,0.08) (1,0.02) };
\addplot [fill=orange] coordinates {(0,0.29) (1,0.25) };
\end{axis}
\end{tikzpicture}
\end{subfigure}
\hspace{-0.4em}  % Reduce space between subfigures
% Second subfigure
\begin{subfigure}[b]{0.249\textwidth}  % Adjusted width
\centering
\begin{tikzpicture}[scale=0.95] 
\begin{axis}[
ybar,
ylabel={\footnotesize F1 top-10},
ylabel style={font=\footnotesize, yshift=-0.5em},
ymin=0,
ymax=0.55,
width=1.2\textwidth,
bar width=1ex,
height=4cm,
xtick=data,
xticklabels={\dbpedia, \lmdb},
xtick pos=bottom,
ytick pos=left,
enlarge x limits=0.4,
yticklabel style={font=\scriptsize, /pgf/number format/fixed},
tick label style={font=\scriptsize},
xticklabel style={rotate=45, anchor=east}
]
\addplot [fill=blue]   coordinates {(0,0.30) (1,0.26) };
\addplot [fill=red]    coordinates {(0,0.14) (1,0.34) };
\addplot [fill=green]  coordinates {(0,0.45) (1,0.24) };
\addplot [fill=yellow] coordinates {(0,0.15) (1,0.14) };
\addplot [fill=orange] coordinates {(0,0.54) (1,0.40) };
\end{axis}
\end{tikzpicture}
\end{subfigure}
\hspace{-0.4em}  % Reduce space between subfigures
% Third subfigure
\begin{subfigure}[b]{0.249\textwidth}  % Adjusted width
\centering
\begin{tikzpicture}[scale=0.95] 
\begin{axis}[
ybar,
ylabel={\footnotesize MAP-5},
ylabel style={font=\footnotesize, yshift=-0.5em},
ymin=0,
ymax=0.22,
width=1.2\textwidth,
bar width=1ex,
height=4cm,
xtick=data,
xticklabels={\dbpedia, \lmdb},
xtick pos=bottom,
ytick pos=left,
enlarge x limits=0.4,
yticklabel style={font=\scriptsize, /pgf/number format/fixed},
tick label style={font=\scriptsize},
xticklabel style={rotate=45, anchor=east}
]
\addplot [fill=blue]   coordinates {(0,0.07) (1,0.06) };
\addplot [fill=red]    coordinates {(0,0.1) (1,0.19) };
\addplot [fill=green]  coordinates {(0,0.13) (1,0.05) };
\addplot [fill=yellow] coordinates {(0,0.04) (1,0.01) };
\addplot [fill=orange] coordinates {(0,0.18) (1,0.13) };
\end{axis}
\end{tikzpicture}
\end{subfigure}
\hspace{-0.4em}  % Reduce space between subfigures
% Fourth subfigure
\begin{subfigure}[b]{0.249\textwidth}  % Adjusted width
\centering
\begin{tikzpicture}[scale=0.95] 
\begin{axis}[
ybar,
ylabel={\footnotesize MAP-10},
ylabel style={font=\footnotesize, yshift=-0.5em},
ymin=0,
ymax=0.4,
width=1.2\textwidth,
bar width=1ex,
height=4cm,
xtick=data,
xticklabels={\dbpedia, \lmdb},
xtick pos=bottom,
ytick pos=left,
enlarge x limits=0.4,
yticklabel style={font=\scriptsize, /pgf/number format/fixed},
tick label style={font=\scriptsize},
xticklabel style={rotate=45, anchor=east}
]
\addplot [fill=blue]   coordinates {(0,0.16) (1,0.13) };
\addplot [fill=red]    coordinates {(0,0.04) (1,0.22) };
\addplot [fill=green]  coordinates {(0,0.29) (1,0.11) };
\addplot [fill=yellow] coordinates {(0,0.06) (1,0.05) };
\addplot [fill=orange] coordinates {(0,0.39) (1,0.23) };
\end{axis}
\end{tikzpicture}
\end{subfigure}
\hspace{-1em}  % Reduce space between subfigures
\caption{F1 score and MAP for frequency statistics on \esbm datasets.}
\label{fig:esbm-stats}
\end{figure}
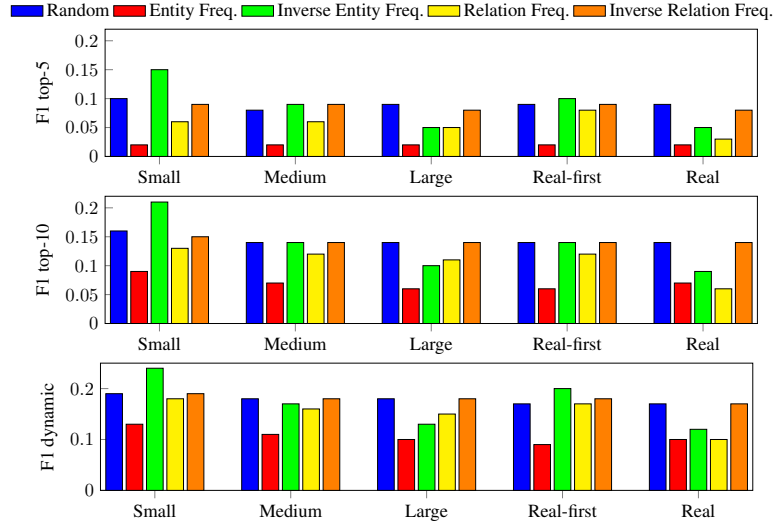
\begin{figure}[htb]
\centering
% First subfigure
\begin{subfigure}[b]{\textwidth}
\centering
\begin{tikzpicture}[scale=0.7] 
\begin{axis}[
hide axis,
xmin=0,
xmax=10,
ymin=0,
ymax=10,
legend columns=5,
legend style={at={(0.5,1)},anchor=north,draw=none,fill=none,legend cell align=left, xshift=2cm},
legend image post style={scale=1},  % Add this line to scale down the legend images
legend entries={Random, Entity Freq., Inverse Entity Freq., Relation Freq., Inverse Relation Freq.}
]
\addlegendimage{fill=blue,area legend}
\addlegendimage{fill=red,area legend}
\addlegendimage{fill=green,area legend}
\addlegendimage{fill=yellow,area legend}
\addlegendimage{fill=orange,area legend}
\end{axis}
\end{tikzpicture}
\begin{tikzpicture}[scale=0.7] 
\begin{axis}[
ybar,
ylabel={F1 top-5},
ymin=0,
ymax=0.22,
width=\textwidth,
bar width=2ex,
height=4cm,  % Adjust the height to fit the vertical arrangement
xtick=data,
xticklabels={Small, Medium, Large , Real-first , Real },
xtick pos=bottom,
ytick pos=left,
enlarge x limits=0.1,
yticklabel style={/pgf/number format/fixed}
]
\addplot [fill=blue]   coordinates {(0,0.10) (1,0.08) (2,0.09) (3,0.09) (4,0.09)};
\addplot [fill=red]    coordinates {(0,0.02) (1,0.02) (2,0.02) (3,0.02) (4,0.02)};
\addplot [fill=green]  coordinates {(0,0.15) (1,0.09) (2,0.05) (3,0.10) (4,0.05)};
\addplot [fill=yellow] coordinates {(0,0.06) (1,0.06) (2,0.05) (3,0.08) (4,0.03)};
\addplot [fill=orange] coordinates {(0,0.09) (1,0.09) (2,0.08) (3,0.09) (4,0.08)};
\end{axis}
\end{tikzpicture}
\end{subfigure}
% Second subfigure
\begin{subfigure}[b]{\textwidth}
\centering
\begin{tikzpicture}[scale=0.7] 
\begin{axis}[
ybar,
ylabel={F1 top-10},
ymin=0,
ymax=0.22,
width=\textwidth,
bar width=2ex,
height=4cm,  % Adjust the height to fit the vertical arrangement
xtick=data,
xticklabels={Small, Medium, Large , Real-first , Real },
xtick pos=bottom,
ytick pos=left,
enlarge x limits=0.1,
yticklabel style={/pgf/number format/fixed}
]
\addplot [fill=blue]   coordinates {(0,0.16) (1,0.14) (2,0.14) (3,0.14) (4,0.14)};
\addplot [fill=red]    coordinates {(0,0.09) (1,0.07) (2,0.06) (3,0.06) (4,0.07)};
\addplot [fill=green]  coordinates {(0,0.21) (1,0.14) (2,0.10) (3,0.14) (4,0.09)};
\addplot [fill=yellow] coordinates {(0,0.13) (1,0.12) (2,0.11) (3,0.12) (4,0.06)};
\addplot [fill=orange] coordinates {(0,0.15) (1,0.14) (2,0.14) (3,0.14) (4,0.14)};
\end{axis}
\end{tikzpicture}
\end{subfigure}
\begin{subfigure}[b]{\textwidth}
\centering
\begin{tikzpicture}[scale=0.7] 
\begin{axis}[
ybar,
ylabel={F1 dynamic},
ymin=0,
ymax=0.25,
width=\textwidth,
bar width=2ex,
height=4cm,  % Adjust the height to fit the vertical arrangement
xtick=data,
xticklabels={Small, Medium, Large , Real-first , Real },
xtick pos=bottom,
ytick pos=left,
enlarge x limits=0.1,
yticklabel style={/pgf/number format/fixed}
]
\addplot [fill=blue]   coordinates {(0,0.19) (1,0.18) (2,0.18) (3,0.17) (4,0.17)};
\addplot [fill=red]    coordinates {(0,0.13) (1,0.11) (2,0.10) (3,0.09) (4,0.10)};
\addplot [fill=green]  coordinates {(0,0.24) (1,0.17) (2,0.13) (3,0.20) (4,0.12)};
\addplot [fill=yellow] coordinates {(0,0.18) (1,0.16) (2,0.15) (3,0.17) (4,0.10)};
\addplot [fill=orange] coordinates {(0,0.19) (1,0.18) (2,0.18) (3,0.18) (4,0.17)};
\end{axis}
\end{tikzpicture}
\end{subfigure}
\caption{F1 for frequency statistics on \fourthdataset.}
\label{fig:f1-wikies}
\vspace{1em}
\end{figure}

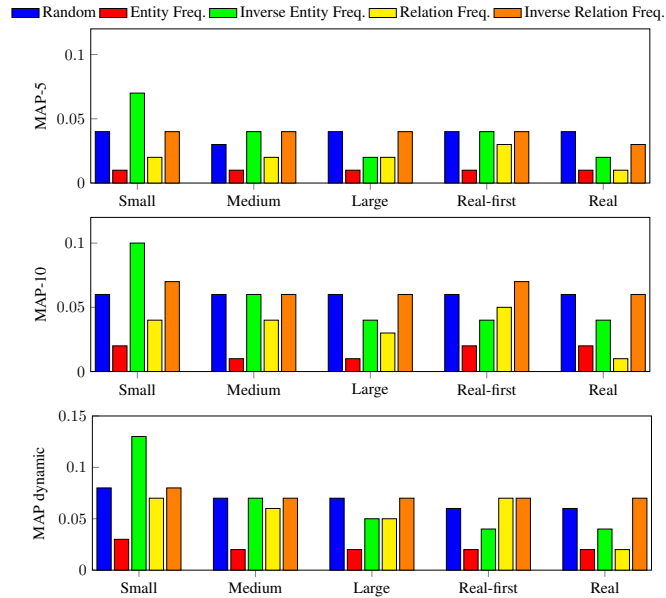
\begin{figure}[htb]
\centering
% First subfigure
\begin{subfigure}[b]{\textwidth}
\centering
\begin{tikzpicture}[scale=0.6] 
\begin{axis}[
hide axis,
xmin=0,
xmax=10,
ymin=0,
ymax=10,
legend columns=5,
legend style={at={(0.5,1)},anchor=north,draw=none,fill=none,legend cell align=left, xshift=2cm},
legend image post style={scale=1},  % Add this line to scale down the legend images
legend entries={Random, Entity Freq., Inverse Entity Freq., Relation Freq., Inverse Relation Freq.}
]
\addlegendimage{fill=blue,area legend}
\addlegendimage{fill=red,area legend}
\addlegendimage{fill=green,area legend}
\addlegendimage{fill=yellow,area legend}
\addlegendimage{fill=orange,area legend}
\end{axis}
\end{tikzpicture}
\begin{tikzpicture}[scale=0.6] 
\begin{axis}[
ybar,
ylabel={MAP-5},
ymin=0,
ymax=0.12,
width=\textwidth,
bar width=2ex,
height=5cm,  % Adjust the height to fit the vertical arrangement
xtick=data,
xticklabels={Small, Medium, Large , Real-first , Real },
xtick pos=bottom,
ytick pos=left,
enlarge x limits=0.1,
yticklabel style={/pgf/number format/fixed}
]
\addplot [fill=blue]   coordinates {(0,0.04) (1,0.03) (2,0.04) (3,0.04) (4,0.04)};
\addplot [fill=red]    coordinates {(0,0.01) (1,0.01) (2,0.01) (3,0.01) (4,0.01)};
\addplot [fill=green]  coordinates {(0,0.07) (1,0.04) (2,0.02) (3,0.04) (4,0.02)};
\addplot [fill=yellow] coordinates {(0,0.02) (1,0.02) (2,0.02) (3,0.03) (4,0.01)};
\addplot [fill=orange] coordinates {(0,0.04) (1,0.04) (2,0.04) (3,0.04) (4,0.03)};
\end{axis}
\end{tikzpicture}
\end{subfigure}
% Second subfigure
\begin{subfigure}[b]{\textwidth}
\centering
\begin{tikzpicture}[scale=0.6] 
\begin{axis}[
ybar,
ylabel={MAP-10},
ymin=0,
ymax=0.12,
width=\textwidth,
bar width=2ex,
height=5cm,  % Adjust the height to fit the vertical arrangement
xtick=data,
xticklabels={Small, Medium, Large , Real-first , Real },
xtick pos=bottom,
ytick pos=left,
enlarge x limits=0.1,
yticklabel style={/pgf/number format/fixed}
]
\addplot [fill=blue]   coordinates {(0,0.06) (1,0.06) (2,0.06) (3,0.06) (4,0.06)};
\addplot [fill=red]    coordinates {(0,0.02) (1,0.01) (2,0.01) (3,0.02) (4,0.02)};
\addplot [fill=green]  coordinates {(0,0.10) (1,0.06) (2,0.04) (3,0.04) (4,0.04)};
\addplot [fill=yellow] coordinates {(0,0.04) (1,0.04) (2,0.03) (3,0.05) (4,0.01)};
\addplot [fill=orange] coordinates {(0,0.07) (1,0.06) (2,0.06) (3,0.07) (4,0.06)};
\end{axis}
\end{tikzpicture}
\end{subfigure}
\begin{subfigure}[b]{\textwidth}
\centering
\begin{tikzpicture}[scale=0.6] 
\begin{axis}[
ybar,
ylabel={MAP dynamic},
ymin=0,
ymax=0.15,
width=\textwidth,
bar width=2ex,
height=5cm,  % Adjust the height to fit the vertical arrangement
xtick=data,
xticklabels={Small, Medium, Large , Real-first , Real },
xtick pos=bottom,
ytick pos=left,
enlarge x limits=0.1,
yticklabel style={/pgf/number format/fixed}
]
\addplot [fill=blue]   coordinates {(0,0.08) (1,0.07) (2,0.07) (3,0.06) (4,0.06)};
\addplot [fill=red]    coordinates {(0,0.03) (1,0.02) (2,0.02) (3,0.02) (4,0.02)};
\addplot [fill=green]  coordinates {(0,0.13) (1,0.07) (2,0.05) (3,0.04) (4,0.04)};
\addplot [fill=yellow] coordinates {(0,0.07) (1,0.06) (2,0.05) (3,0.07) (4,0.02)};
\addplot [fill=orange] coordinates {(0,0.08) (1,0.07) (2,0.07) (3,0.07) (4,0.07)};
\end{axis}
\end{tikzpicture}
\end{subfigure}
\caption{MAP for frequency statistics on \fourthdataset.}
\label{fig:map-wikies}
\vspace{1em}
\end{figure}

%\subsection{Entity summarization methods}
\dm{We evaluate the performance of different entity summarization methods on our benchmark, and provide all implementations in the \ourdataset GitHub repository.}

\dm{\begin{itemize}[leftmargin=*]
    \item \textbf{PageRank}~\cite{ma2008bringing} ranks nodes in a graph based on the structure of incoming links, with the idea that more important nodes are likely to receive more links from other nodes.
    \item \textbf{RELIN}~\cite{cheng2011relin} is a weighted PageRank algorithm that evaluates the relevance of triples within a graph structure. We have re-implemented this model according to the specifications in the referenced paper. On our smaller dataset version, RELIN takes approximately 6 hours to compute all summaries.
    \item\textbf{LinkSum}~\cite{thalhammer2016linksum} is a two-step, relevance-centric method that combines PageRank with an adaptation of the Backlink algorithm to identify relevant connected entities. We have re-implemented it according to the paper. The LinkSum method initially takes 10 hours to compute the backlinks for each node in the small version of our dataset. By parallelizing the implementation, we reduced this to one hour. Additionally, the Backlink algorithm itself initially takes 100 minutes, but with parallelization, this was reduced to 10 minutes for the small version of our dataset.
\end{itemize}}

\dm{Due to the inefficiency of the methods, we use a smaller version of \ourdataset for evaluation. The results in Table~\ref{exp:baseline-results} show that LinkSum outperforms both RELIN and PageRank. These findings suggest that models capable of exploiting the graph structure while handling large-scale datasets and maintaining high accuracy in entity summarization are valuable for such real-world KGs, such as \ourdataset.}

\begin{table}[ht]
\footnotesize
    \centering
    \begin{tabular}{llcc|cc|cc}
    \toprule
    & & \multicolumn{2}{c}{topK = 5} & \multicolumn{2}{c}{topK = 10} & \multicolumn{2}{c}{ Dynamic} \\
    \cmidrule(lr){3-4} \cmidrule(lr){5-6} \cmidrule(lr){7-8}
    Model & Dataset & F-Score & MAP & F-Score & MAP &F-Score & MAP  \\
    \midrule
    PageRank  & \firstdataset & 0.024 & 0.01 & 0.081 & 0.02 & 0.175 & 0.046\\
              & \seconddataset & 0.003 & 0.001 & 0.041 & 0.005 & 0.146 & 0.028\\
              & \thirddataset & 0.060 & 0.02 & 0.169 & 0.049 & 0.288 & 0.109\\
              & \fourthdataset & 0.032 & 0.01 & 0.093 & 0.024 & 0.145 & 0.036\\
    RELIN     & \firstdataset & 0.093 & 0.035 & 0.148 & 0.054 & 0.208 & 0.080\\
              & \seconddataset & 0.071 & 0.023 & 0.127 & 0.038 & 0.209 & 0.068\\
              & \thirddataset & 0.125 & 0.053 & 0.200& 0.086 & 0.273 & 0.127\\
              & \fourthdataset & 0.111 & 0.050 & 0.179 & 0.081 & 0.219 & 0.095\\
    LinkSum   & \firstdataset & 0.184 & 0.080 & 0.239 & 0.109 & 0.225 & 0.127\\
              & \seconddataset & 0.119 & 0.048 & 0.152 & 0.060 & 0.135 & 0.068\\
              & \thirddataset & 0.249 & 0.127 & 0.347 & 0.190 & 0.350 & 0.242\\
              & \fourthdataset & 0.195 & 0.097 & 0.236 & 0.127 & 0.213 & 0.136\\
    \bottomrule
    
    \end{tabular}
  \caption{Performance comparison of entity summarization models on the small version of \ourdataset. The models are evaluated with different topK values (5 and 10) and a dynamic setting.}
\label{exp:baseline-results}
\end{table}

\section{Conclusion}
\label{sec:conclusions}

\dm{We introduce \ourdataset (Wiki Entity Summarization Benchmark), a benchmark for KG entity summarization which provides a scalable dataset generator that eschews the need for costly human annotation.}
\dm{\ourdataset uses Wikipedia abstracts for automatic summary generation, ensuring contextually rich and unbiased summaries. It preserves the complexity and integrity of real-world KGs through a random walk sampling method that captures the structure of entities down to their second-hop neighborhoods.}
%\sj{However, our approach has some limitations. There is a minor node frequency bias in small datasets, which diminishes with larger sizes. Seed nodes should be Wikipedia page IDs with abstracts to ensure quality summaries. We also enforced a minimum summary size, potentially excluding some shorter entries. }
\dm{Empirical evaluations demonstrate that \ourdataset provides high-quality large-scale datasets for entity summarization tasks, and that it captures the complexities of knowledge graphs in terms of topology, making it a valuable resource for evaluating and improving entity summarization algorithms. }

% Future research opportunities with \ourdataset include:

% \begin{itemize}[leftmargin=*, noitemsep]

%     \item\dm{\textbf{Enhanced Model Evaluation:} \ourdataset can be utilized to assess the influence of different summarization techniques and model architectures on entity summarization performance.}
%     \item\dm{\textbf{Scalability Testing:}  We encourage researchers to use \ourdataset to explore the scalability of summarization algorithms, given its ability to generate large and complex datasets.}
%     \item\dm{\textbf{Integration of Graph and Textual Data: } Incorporating textual information from Wikipedia abstracts along with graph structures for entity summarization tasks can significantly enhance model accuracy. Additionally, this integration can facilitate the development of advanced methods for few-shot summarization and other emerging research areas.}
% \end{itemize}

% Overall, we believe that \ourdataset will significantly advance research in entity summarization and, consequently, impact various fields of artificial intelligence.}

%\section{Limitations and future work}
% \cite{akbik2018contextual2}

\newpage

%\small

%\bibliographystyle{plain}
%\bibliography{references}

%%%%%%%%%%%%%%%%%%%%%%%%%%%%%%%%%%%%%%%%%%%%%%%%%%%%%%%%%%%%

\appendix
\section{Appendix}
\label{sec:appendix}

\begin{table}[H]
\small % Reduce font size
\caption{Generated Datasets in terms of number of entities $|\V|$, triples $|\E|$, ground-truth summaries, density as $\displaystyle|\E|/\binom{|\V|}{2}$, graph connectivity, number of components, sampling method to select the entities and the subgraph, minimum and maximum node degree and, running time.}
\begin{subtable}{\linewidth}
\centering
\caption{Small Datasets}
\begin{tabularx}{\linewidth}{lcccc} % Adjusted column definition
\toprule
Metric & \firstdataset & \seconddataset & \thirddataset & \fourthdataset \\ 
\midrule
Entities ($|\V|$) & 85\,346 & 70\,753 & 79\,825 & 79\,926 \\ 
Relations ($|\E|$) & 136\,950 & 126\,915 & 125\,912 & 123\,193 \\ 
Target Entities & 494 & 493 & 493 & 468 \\ 
Density & 0.000018 & 0.000018 & 0.000019 & 0.000019 \\ 
Sampling method & Random Walk & Random Walk & Random Walk & Random Walk \\ 
Connected-graph & Yes & Yes & Yes & Yes \\ 
Num-comp & 1 & 1 & 1 & 1 \\ 
Min Degree & 1 & 1 & 1 & 1 \\ 
Max Degree & 2172 & 3005 & 2060 & 3142 \\ 
Run-time (seconds) & 91.934 & 118.014 & 126.119 & 177.63 \\ 
\bottomrule
\end{tabularx}
\end{subtable}

\vspace{0.5cm} % Space between tables

\begin{subtable}{\linewidth}
\centering
\caption{Medium Datasets}
\begin{tabularx}{\linewidth}{lcccc} % Adjusted column definition
\toprule
Metric &  \firstdataset & \seconddataset & \thirddataset & \fourthdataset \\ 
\midrule
Entities ($|\V|$) & 128\,061 & 101\,529 & 119\,305 & 122\,728 \\ 
Relations ($|\E|$) & 220\,263 & 196\,061 & 198\,663 & 196\,838 \\ 
Target Entities & 494 & 493 & 493 & 468 \\ 
Density & 0.000013 & 0.000019 & 0.000014 & 0.000013 \\ 
Sampling method & Random Walk & Random Walk & Random Walk & Random Walk \\ 
Connected-graph & Yes & Yes & Yes & Yes \\ 
Num-comp & 1 & 1 & 1 & 1 \\ 
Min Degree & 1 & 1 & 1 & 1 \\ 
Max Degree & 3726 & 5124 & 3445 & 5282 \\ 
Run-time (seconds) & 155.36 & 196.413 & 208.157 & 301.718 \\ 
\bottomrule
\end{tabularx}
\end{subtable}

\vspace{0.5cm} % Space between tables

\begin{subtable}{\linewidth}
\centering
\caption{Large Datasets}
\begin{tabularx}{\linewidth}{lcccc} % Adjusted column definition
\toprule
Metric & \firstdataset & \seconddataset & \thirddataset & \fourthdataset \\ 
\midrule
Entities ($|\V|$) & 239\,491 & 185\,098 & 230\,442 & 248\,012 \\ 
Relations ($|\E|$) & 466\,905 & 397\,546 & 412\,766 & 413\,895 \\ 
Target Entities & 494 & 493 & 493 & 468 \\ 
Density & 0.000008 & 0.00001 & 0.000008 & 0.000007 \\ 
Sampling method & Random Walk & Random Walk & Random Walk & Random Walk \\ 
Connected-graph & Yes & Yes & Yes & Yes \\ 
Num-comp & 1 & 1 & 1 & 1 \\ 
Min Degree & 1 & 1 & 1 & 1 \\ 
Max Degree & 8599 & 12189 & 7741 & 12939 \\ 
Run-time (seconds) & 353.113 & 475.679 & 489.409 & 768.99 \\ 
\bottomrule
\end{tabularx}
\end{subtable}
\label{tbl:combined-dataset}
\end{table}

\subsection{Additional \ourdataset details }

\begin{itemize}

    %\item \textbf{Purpose and motivation behind the \ourdataset Generator and Benchmark:} The motivation behind this dataset generator is to make it available for researchers so that they can generate datasets quickly without the cost of human annotation. We create an easy-to-use executable generator, in which users can create connected graphs by passing their desired seed nodes.Furthermore, we generate some datasets because there has not been a large dataset in this field, and consequently, scalable models have not been developed in this area. To address this issue, we generate large datasets alongside their medium and small versions so that researchers can use them. By running the current models using our generated datasets, we can pinpoint that the current models are not scalable, which will guide researchers in developing more scalable models.
    %Unlike previous benchmarks, we were interested in generating datasets that follow the patterns of real-world graphs and do not have a bias towards any metric that does not exist in real data. In Section~\ref{sec:evaluation}, we study our dataset for different biases and compare it to the \esbm benchmark, which is the most recent entity summarization benchmark to date.

    \item \textbf{Dataset and Metadata:} The dataset is available at \url{https://github.com/msorkhpar/wiki-entity-summarization}. We generate four datasets in three sizes: small, medium, and large. Each size has a train-test-validation split. In Table~\ref{tab:seed_nodes}, we provide information on the proportion of seed nodes in each of the datasets. Moreover, Table~\ref{tbl:combined-dataset} provides detailed information such as the number of entities $|\V|$, triples $|\E|$, ground-truth summaries, density, graph connectivity, number of components, sampling method used to select the entities and the subgraph, minimum and maximum node degree, and running time for each of the datasets.
    Moreover, metadata is also in the same github repository. 
    
    \item \textbf{\ourdataset toolkits:} We offer a comprehensive toolkit designed to facilitate working with our datasets. This toolkit includes features for downloading, loading, and manipulating pre-generated graph datasets. You can access our toolkit at \url{https://pypi.org/project/wikes-toolkit/}
    
    \item \textbf{Dataset Formats:} We generate our dataset in CSV format. The entity files are formatted according to Table~\ref{tbl:appendix:entities}. Additionally, we provide a file that contains the categories of the target entities, as detailed in Table~\ref{tbl:appendix:groundtruth}. The predicate files, described in Table~\ref{tbl:appendix:predicate}, contain predicates along with their corresponding labels and descriptions. The triple file, presented in Table~\ref{tbl:appendix:triples}, includes the subject, object, and predicate IDs of the nodes (Wikidata items) and edges (Wikidata predicates). The ground-truth file, shown in Table~\ref{tbl:appendix:groundtruth}, contains the subject, object, and predicate. Predicates in groun-truth entities are also marked as ground-truth.  Moreover, we provide the graph version of our dataset in GraphML and PKL formats in our release \url{https://github.com/msorkhpar/wiki-entity-summarization/releases/tag/1.0.5}. 
    
    \item \textbf{URL to metadata record:} Since we have several datasets, in our GitHub Repository, we provide the Crossiant URL metadata. The metadata for \firstdataset small dataset is \url{https://github.com/msorkhpar/wiki-entity-summarization/releases/download/1.0.5/WikiLitArt-m.json}. The metadata format is consistent across all datasets. The metadata files are included alongside the datasets in the GitHub release.

    \item \textbf{Preprocessing URL:} You can find our preprocessing code for cleaning and preparing Wikipedia and Wikidata at the following link: \url{https://github.com/msorkhpar/wiki-entity-summarization-preprocessor}. 
    Access and loading of preprocessed:
    \begin{itemize}
    \item Neo4j database: \url{https://github.com/msorkhpar/wiki-entity-summarization-preprocessor/releases/tag/Neo4j-1.0.0}.
    \item PostgreSQL: \url{https://github.com/msorkhpar/wiki-entity-summarization-preprocessor/releases/tag/PostgreSQL-1.0.0}.
    \end{itemize}
    
    \item \textbf{Authors responsibility statement and License:} The authors are held responsible for copyright infringement, but assume no responsibility or liability for any misuse of the data. This project is licensed under the CC BY 4.0 License. See here \url{https://github.com/msorkhpar/wiki-entity-summarization/blob/main/LICENSE}
    \item \textbf{\ourdataset Generator Code: } The code for running the \ourdataset generator is available in the GitHub repository at \url{https://github.com/msorkhpar/wiki-entity-summarization}. The code allows to generate the same datasets  in the paper or create your own custom datasets.
    \item \textbf{Maintenance and Long Term Preservation } The authors of \ourdataset are dedicated to the ongoing maintenance and preservation of this dataset. This includes tracking and resolving issues identified by the community post-release. We will closely monitor user feedback through the GitHub issue tracker. The data is hosted on GitHub, ensuring reliable and stable storage. % Depending on usage, we may transition to archival storage for long-term preservation.

    \item \textbf{Intended users:} The intended users are NLP and knowledge graph researchers who wish to generate summaries using the textual information of the entities (nodes) in knowledge graphs. The suitable use case for this dataset is evaluating entity summarization models to determine their ability to accurately detect summaries. 
\end{itemize}

\begin{table}[h]
    \centering
    \begin{tabular}{|l|l|l|}
        \hline
        Field & Description & Datatype \\ \hline
        id & Incremental integer starting by zero & int \\ \hline
        entity & Wikidata qid, e.g. `Q76` & string \\ \hline
        wikidata\_label & Wikidata label (nullable) & string \\ \hline
        wikidata\_desc & Wikidata description (nullable) & string \\ \hline
        wikipedia\_title & Wikipedia title (nullable) & string \\ \hline
        wikipedia\_id & Wikipedia page id (nullable) & long \\ \hline
    \end{tabular}
    \caption{{\texttt{\{variant\}-\{size\}-\{dataset\_type\}-entities.csv}} file contains entities. An entity is a Wikidata item (node) in our dataset. variant\_index\ refers to the dataset id (detailed information is in our Github). }
    \label{tbl:appendix:entities}
\end{table}

\begin{table}[h]
    \centering
    \begin{tabular}{|l|l|l|}
        \hline
        Field & Description & Datatype \\ \hline
        entity & id key in Table~\ref{tbl:appendix:entities} & int \\ \hline
        category & category & string \\ \hline
    \end{tabular}
    \caption{{\texttt{\{variant\}-\{size\}-\{dataset\_type\}-root-entities.csv}} contains root entities. A root entity is a seed node described previously.variant\_index\ refers to the dataset id (detailed infomation is in our Github).}
    \label{tbl:appendix:root-entities}
\end{table}

\begin{table}[h]
    \centering
    \begin{tabular}{|l|l|l|}
        \hline
        Field & Description & Datatype \\ \hline
        id & Incremental integer starting by zero & int \\ \hline
        predicate & Wikidata Property id, e.g. `P121` & string \\ \hline
        predicate\_label & Wikidata Property label (nullable) & string \\ \hline
        predicate\_desc & Wikidata Property description (nullable) & string \\ \hline
    \end{tabular}
    \caption{{\texttt{\{variant\}-\{size\}-\{dataset\_type\}-predicates.csv}} contains predicates. A predicate is a Wikidata property or a describing a connection. variant\_index\ refers to the dataset id (detailed information is in our Github).}
    \label{tbl:appendix:predicate}
\end{table}

\begin{table}[h]
    \centering
    \begin{tabular}{|l|l|l|}
        \hline
        Field & Description & Datatype \\ \hline
        subject & id key in Table~\ref{tbl:appendix:entities} & int \\ \hline
        predicate & id key in Table~\ref{tbl:appendix:predicate} & int \\ \hline
        object & id key in Table~\ref{tbl:appendix:entities} & int \\ \hline
    \end{tabular}
    \caption{{\texttt{\{variant\}-\{size\}-\{dataset\_type\}-triples.csv}} contains triples. A triple is an edge between two entities with a predicate. variant\_index\ refers to the dataset id (detailed information is in our Github).}
    \label{tbl:appendix:triples}
\end{table}

\begin{table}[h]
    \centering
    \begin{tabular}{|l|l|l|}
        \hline
        Field & Description & Datatype \\ \hline
        root\_entity & entity in Table~\ref{tbl:appendix:entities} & int \\ \hline
        subject & id key in Table~\ref{tbl:appendix:entities} & int \\ \hline
        predicate & id key in Table~\ref{tbl:appendix:predicate} & int \\ \hline
        object & id key in Table~\ref{tbl:appendix:entities} & int \\ \hline
    \end{tabular}
    \caption{{\texttt{\{variant\}-\{size\}-\{dataset\_type\}-ground-truths.csv}} contains ground truth triples. A ground truth triple is an edge marked as a summary for a root entity.}
    \label{tbl:appendix:groundtruth}
\end{table}

\begin{table}[H]
\centering
\begin{tabularx}{\textwidth}{|l|X|}
\hline
\rowcolor{lightgray}\textbf{Dataset} & \textbf{Seed Nodes Categories} \\ \hline
\multirow{4}{*}{\firstdataset} & \textbf{Entire graph:} actor=150, composer=35, film=41, novelist=24, painter=59, poet=39, screenwriter=17, singer=72, writer=57 \\ \cline{2-2}
 & \textbf{Train:} actor=105, composer=24, film=29, novelist=17, painter=42, poet=27, screenwriter=12, singer=50, writer=40 \\ \cline{2-2}
 & \textbf{Val:} actor=23, composer=5, film=6, novelist=4, painter=9, poet=6, screenwriter=2, singer=11, writer=8 \\ \cline{2-2}
 & \textbf{Test:} actor=22, composer=6, film=6, novelist=3, painter=8, poet=6, screenwriter=3, singer=11, writer=9 \\ \hline
\multirow{4}{*}{\seconddataset} & \textbf{Entire graph:} actor=405, film=88 \\ \cline{2-2}
 & \textbf{Train:} actor=284, film=61 \\ \cline{2-2}
 & \textbf{Val:} actor=59, film=14 \\ \cline{2-2}
 & \textbf{Test:} actor=62, film=13 \\ \hline
\multirow{4}{*}{{\thirddataset} } & \textbf{Entire graph:} actor=58, football=156, journalist=14, lawyer=16, painter=23, player=25, politician=125, singer=27, sport=21, writer=28 \\ \cline{2-2}
 & \textbf{Train:} actor=41, football=109, journalist=10, lawyer=11, painter=16, player=17, politician=87, singer=19, sport=15, writer=20 \\ \cline{2-2}
 & \textbf{Val:} actor=9, football=23, journalist=2, lawyer=3, painter=3, player=4, politician=19, singer=4, sport=3, writer=4 \\ \cline{2-2}
 & \textbf{Test:} actor=8, football=24, journalist=2, lawyer=2, painter=4, player=4, politician=19, singer=4, sport=3, writer=4 \\ \hline
\multirow{4}{*}{\fourthdataset} & \textbf{Entire graph:} actor=141, athletic=25, football=24, journalist=16, painter=16, player=32, politician=81, singer=69, sport=18, writer=46 \\ \cline{2-2}
 & \textbf{Train:} actor=98, athletic=18, football=17, journalist=9, painter=13, player=22, politician=57, singer=48, sport=14, writer=34 \\ \cline{2-2}
 & \textbf{Val:} actor=21, athletic=4, football=3, journalist=4, painter=1, player=5, politician=13, singer=11, sport=1, writer=5 \\ \cline{2-2}
 & \textbf{Test:} actor=22, athletic=3, football=4, journalist=3, painter=2, player=5, politician=11, singer=10, sport=3, writer=7 \\ \hline
\end{tabularx}
\caption{Seed nodes categories for each dataset. "Entire graph" refers to using the seed nodes and generating the data without train-test-val splits. In train-test-val, each of the datasets is a single weakly connected graph.}

\label{tab:seed_nodes}
\end{table}

\subsection{Parameters for Running the \ourdataset Generator}
Table~\ref{table:parameters} shows the parameters required for running the \ourdataset Generator. The table provides a description of the parameters and their default values, where applicable. A detailed explanation on how to run the generator can be found in our GitHub repository.

\begin{table}[h!]
\centering
\begin{tabular}{|l|p{6cm}|p{5cm}|}
\hline
\textbf{Parameter} & \textbf{Description} & \textbf{Default Value} \\ \hline
min\_valid\_summary\_edges & Minimum number of valid summaries for a seed node & 5 \\ \hline
random\_walk\_depth\_len & Depth length of random walks (number of nodes in each random walk) & 3 \\ \hline
bridges\_number & Number of connecting path bridges between components & 5 \\ \hline
max\_threads & Maximum number of threads & 4 \\ \hline
%output\_path & Path to save output data & N/A \\ \hline
%db\_name & Database name & N/A \\ \hline
%db\_user & Database user & N/A \\ \hline
%db\_password & Database password & N/A \\ \hline
%db\_host & Database host & N/A \\ \hline
%db\_port & Database port & N/A \\ \hline
%neo4j\_user & Neo4j user & N/A \\ \hline
%neo4j\_password & Neo4j password & N/A \\ \hline
%neo4j\_host & Neo4j host & N/A \\ \hline
%neo4j\_port & Neo4j port & N/A \\ \hline
%dataset\_name & The name of the dataset to process & N/A \\ \hline
min\_random\_walk\_number & Minimum number of random walks for each seed node, explained & 100 for small, 150 for medium, and 300 for large \\ \hline
max\_random\_walk\_number & Maximum number of random walks for each seed node & 300 for small, 600 for medium, and 1800 for large \\ \hline
%seed\_node\_ids & Seed node ids in comma separated format & N/A \\ \hline
%categories & Seed node categories in comma separated format & N/A \\ \hline
\end{tabular}
\caption{Parameters for Running \ourdataset Generator}
\label{table:parameters}
\end{table}

\subsection{Technologies}

Table~\ref{app:technology} presents the versions of the technologies and configurations that we use in this work.
\begin{table}[H]
\small % Reduce font size
\centering
\caption{Technology and Configuration Details for Daatset Generations}
\begin{subtable}{\linewidth}
\centering
\caption{Technologies Used: Software Versions and Data Sources}
\begin{tabularx}{\linewidth}{lX} % Adjusted column definition
\toprule
Technology & Version/Details \\ 
\midrule
Java & Version 21 \\ 
Spring Boot & Version 3 \\ 
Docker & Version 24.0.8 \\ 
Python & Version 3.10 \\ 
PostgreSQL & Version 16.3 \\ 
Neo4j & Version 5.20.0-community \\ 
Wikipedia XML Article Dump Files & Published by Wikimedia on 2023/05/01 \\ 
Wikidata XML Article Dump Files & Published by Wikimedia on 2023/05/01 \\ 
\bottomrule
\end{tabularx}
\end{subtable}

\vspace{1em}

\begin{subtable}{\linewidth}
\centering
\caption{Hardware- Spec: Specifications of the AWS EC2 Instance (r5a.4xlarge) }
\begin{tabularx}{\linewidth}{lX} % Adjusted column definition
\toprule
Specification & Details \\ 
\midrule
vCPU & 16 (AMD EPYC 7571, 16 MiB cache, 2.5 GHz) \\ 
Memory & 128 GB (DDR4, 2667 MT/s) \\ 
Storage & 500 GB (EBS, 2880 Max Bandwidth) \\ 
\bottomrule
\end{tabularx}
\end{subtable}
\label{app:technology}
\end{table}
%Include extra information in the appendix. This section will often be part of the supplemental material. Please see the call on the NeurIPS website for links to additional guides on dataset publication.

\section{Datasheet}
\subsection{Motivation}
\begin{enumerate}
    \item \textit{For what purpose was the dataset created?} The motivation behind this dataset generator is to foster research on entity summarization and provide  a tool to generate arbitrary-size datasets eschewing the cost of human annotation. We create an easy-to-use executable generator, in which users can create connected graphs by passing their desired seed nodes.
    Furthermore, we provide some curated datasets because there has not been a large dataset in this field, and consequently, scalable models have not been developed in this area. To address this issue, we generate large datasets alongside their medium and small versions for research. By running the current models using our generated datasets, we can pinpoint scalability issues of current models and nurture research in this area.
    Unlike previous benchmarks, we were interested in generating datasets that follow the patterns of real-world graphs and do not have a bias towards any metric that does not exist in real data. In Section~\ref{sec:evaluation}, we study our dataset for different biases and compare it to the \esbm benchmark, which is the most recent entity summarization benchmark to date.
    \item \textit{Who created the dataset and on behalf of which entity?} The dataset is fully created by the authors of this paper.
    %\item Who funded the creation of the dataset? Atefeh Moradan is supported by the Innovationsfonden Denmark under the Grand Solutions project Hospital@Night.
\end{enumerate}

\subsection{Distribution}
\begin{enumerate}
    \item \textit{Will the dataset be distributed to third parties outside of the entity (e.g., company, institution, organization) on behalf of which the dataset was created?} Yes, the dataset is open and fully accessible.
    \item \textit{How will the dataset be distributed (e.g., tarball on website, API, GitHub)?} The \ourdataset generator, \ourdataset benchmark and, the code for developing the baselines will be distributed through our GitHub repository.
    
    \item \textit{Have any third parties imposed IP-based or other restrictions on the data associated with the instances?} No.
    \item \textit{Do any export controls or other regulatory restrictions apply to the dataset or to individual instances?} No.
\end{enumerate}

\subsection{Maintenance}
\begin{enumerate}
    \item \textit{Who will be supporting/hosting/maintaining the dataset?}  The authors of the paper.

    \item \textit{How can the owner/curator/manager of the dataset be contacted (e.g., email address)?} The owner/curator/manager(s) of the dataset can be contacted through the following emails: Mohammad Sorkhpar (\texttt{msorkhpar@sycamores.indstate.edu}), Saeedeh Javadi (\texttt{saeedeh.javadi@studenti.polito.it}), and Atefeh Moradan (\texttt{atefeh.moradan@cs.au.dk}).
    
    \item \textit{Is there an erratum?} No, but the users can submit github issues or contact the authors. If errors are found in the future, we will release errata on the main web page for the dataset (\url{https://github.com/msorkhpar/wiki-entity-summarization}).
    
    \item \textit{Will the dataset be updated (e.g., to correct labeling errors, add new instances, delete instances)?} Yes, the datasets will be updated whenever necessary to ensure accuracy, with announcements made accordingly. These updates will be posted on the main webpage for the dataset (\url{https://github.com/msorkhpar/wiki-entity-summarization}).

    \item \textit{If the dataset relates to people, are there applicable limits on the retention of the data associated with the instances (e.g., were the individuals in question told that their data would be retained for a fixed period of time and then deleted?}) N/A.
    
    \item \textit{Will older versions of the dataset continue to be supported/hosted/maintained?} Yes, older versions of the dataset will continue to be maintained and hosted
    \item \textit{If others want to extend/augment/build on/contribute to the dataset, is there a mechanism for them to do so?} Yes, as described above.
\end{enumerate}

\subsection{Composition}

\begin{enumerate}
    \item \textit{What do the instances that comprise the dataset represent (e.g., documents, photos, people, countries)?} Each instance is a document of a Wikidata item, containing some of its immediate neighbours sampled from the Wikidata Knowledge Graph.
    
    \item \textit{How many instances are there in total (of each type, if appropriate)? }See Section~\ref{sec:dataset-info}
    
    \item \textit{Does the dataset contain all possible instances or is it a sample of instances from a larger set?} It is a sample of instance of a larger dataset from Wikipedia and Wikidata. See section ~\ref{algo:randwalk} 
    \item \textit{Is there a label or target associated with each instance?} Yes, each target entity includes target variables.
    
    \item \textit{Is any information missing from individual instances?} No, we include all the information from the intersection of Wikipedia with Wikidata.
    \item \textit{Are there recommended data splits (e.g., training, development/validation, testing)?} We have generate a train-validation-test for each of the graphs that we generate. See section~\ref{sec:dataset-info}
    \item \textit{Are there any errors, sources of noise, or redundancies in the dataset?} Not that we are aware of. 
    
    \item Is the dataset self-contained, or does it link to or otherwise rely on external resources (e.g., websites, tweets, other datasets)? The dataset is self-contained.
    
    \item \textit{Does the dataset contain data that might be considered confidential?} No.
    
    \item \textit{Does the dataset contain data that, if viewed directly, might be offensive, insulting, threatening, or might otherwise cause anxiety?} No.
\end{enumerate}

\subsection{Collection Process}
\begin{enumerate}
    \item \textit{How was the data associated with each instance acquired?} Each instance in this dataset is acquired from raw XML data dumps published by the Wikimedia Foundation on 2023/05/01. The data is then cleaned and processed by our pre-processing module to extract entities, their relations, and their associated summaries. For more information on the summary annotation process, please refer to Section \ref{sec:preprocess}, ``Summary Annotation.''
    %The data associated with each instance is acquired from a series of simulations of a global climate model called E3SM-MMF. References for E3SM-MMF are provided in Section 3 of the main text.
    \item \textit{What mechanisms or procedures were used to collect the data (e.g., hardware apparatus or sensor, manual human curation, software program, software API)?} We utilized an AWS EC2 r5a.4xlarge machine to clean the data and build the datasets using a fully automated pipeline. However, Wikipedia admins and contributors can be considered indirect annotators of this dataset.
    %We used many NVIDIA A100 GPU nodes in a high-performance computing cluster called Perlmutter (operated by the U.S. Department of Energy) to run the E3SM-MMF simulations.
    \item \textit{Who was involved in the data collection process (e.g., students, crowdworkers, contractors) and how were they compensated (e.g., how much were crowdworkers paid)?} Regular employees and two master students were involved in the data collection process. However, the annotation process was fully automated, utilizing publicly available information.
    %Regular employees (e.g., scientists and postdocs) at UC Irvine, LLNL, and SNL were involved in the data collection process. No crowdworkers were involved during the data collection process.
    \item \textit{Does the dataset relate to people?} No.
    
    \item \textit{Did you collect the data from the individuals in question directly, or obtain it via third parties or other sources (e.g., websites)?} We sample the dataset from Wikimedia organization dump files on 2023/05/01.

    %We obtained the dataset from computer simulations of Earth’s climate.
\end{enumerate}

\subsection{Uses}
\begin{enumerate}
    \item \textit{Has the dataset been used for any tasks already?}  We have solely used the dataset for training and evaluating entity summarization baselines in the paper. 
    %No, this dataset has not been used for any tasks yet.
    \item \textit{What (other) tasks could the dataset be used for?} Even though, the main target of this dataset is the Entity summarization task it might also be used for link prediction tasks in knowledge graphs.
    %Please refer to Section 5 in the main manuscript for other applications.
    \item \textit{Is there anything about the composition of the dataset or the way it was collected and preprocessed/cleaned/labeled that might impact future uses? }
    This dataset is based on cleaned and annotated pre-processed information from Wikimedia published dump files on 2023/05/01. Using this pre-processed dump, one can generate a new dataset based on their desired seed node. However, to generate an updated version of the Wikidata knowledge graph, the pre-processing dumps should be regenerated from updated dump files. The instructions and code for this task are publicly available in our preprocessing URL. 
    %The current composition of the datasets is self-sufficient to build a climate emulator. However, it misses some extra variables, which are not essential for such climate emulators but necessary to strictly enforce physical constraints (see Section 4.5 of the main text). We plan to include these extra variables in the next release. Any changes in the next release and updates to user guidelines will be documented and shared through the dataset webpage (\url{https://leap-stc.github.io/ClimSim}).
    \item \textit{Are there tasks for which the dataset should not be used?} No.
\end{enumerate}

\newpage
\section{Experiments}

We include the experiments in Section~\ref{sec:evaluation} for all of our datasets below. 

% \begin{figure}[htb]
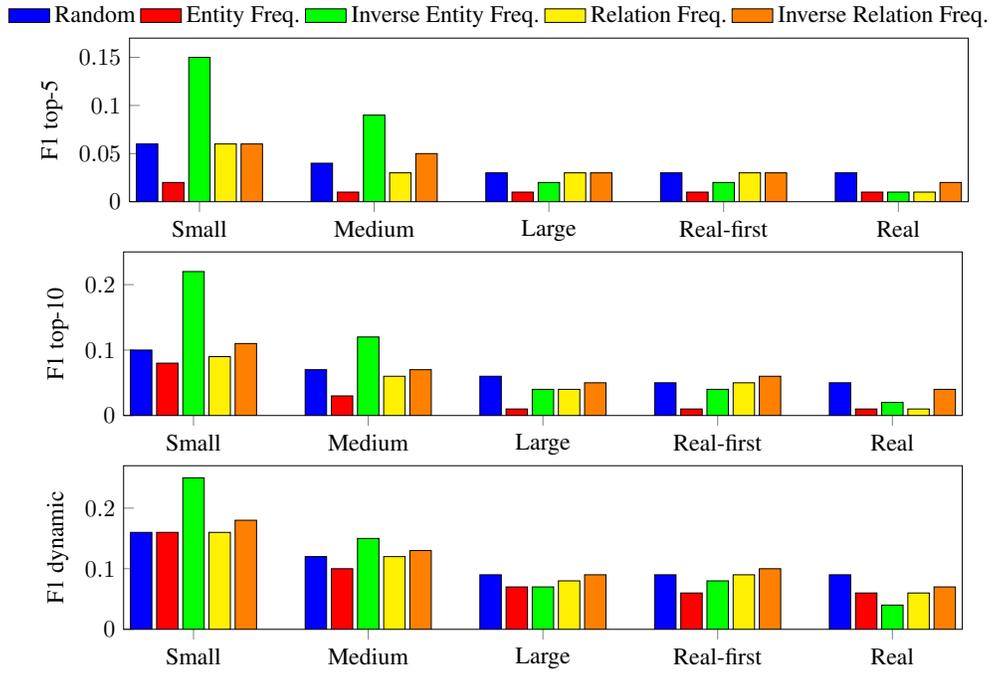
\begin{figure}[H]
\centering
% First subfigure
\begin{subfigure}[b]{\textwidth}
\centering
\begin{tikzpicture}[scale=0.9] 
\begin{axis}[
hide axis,
xmin=0,
xmax=10,
ymin=0,
ymax=10,
legend columns=5,
legend style={at={(0.5,1)},anchor=north,draw=none,fill=none,legend cell align=left, xshift=2cm},
legend image post style={scale=1},  % Add this line to scale down the legend images
legend entries={Random, Entity Freq., Inverse Entity Freq., Relation Freq., Inverse Relation Freq.}
]
\addlegendimage{fill=blue,area legend}
\addlegendimage{fill=red,area legend}
\addlegendimage{fill=green,area legend}
\addlegendimage{fill=yellow,area legend}
\addlegendimage{fill=orange,area legend}
\end{axis}
\end{tikzpicture}
\begin{tikzpicture}[scale=0.9] 
\begin{axis}[
ybar,
ylabel={F1 top-5},
ymin=0,
ymax=0.17,
width=\textwidth,
bar width=2ex,
height=4cm,  % Adjust the height to fit the vertical arrangement
xtick=data,
xticklabels={Small, Medium, Large , Real-first , Real },
xtick pos=bottom,
ytick pos=left,
enlarge x limits=0.1,
yticklabel style={/pgf/number format/fixed}
]
\addplot [fill=blue]   coordinates {(0,0.06) (1,0.04) (2,0.03) (3,0.03) (4,0.03)};
\addplot [fill=red]    coordinates {(0,0.02) (1,0.01) (2,0.01) (3,0.01) (4,0.01)};
\addplot [fill=green]  coordinates {(0,0.15) (1,0.09) (2,0.02) (3,0.02) (4,0.01)};
\addplot [fill=yellow] coordinates {(0,0.06) (1,0.03) (2,0.03) (3,0.03) (4,0.01)};
\addplot [fill=orange] coordinates {(0,0.06) (1,0.05) (2,0.03) (3,0.03) (4,0.02)};
\end{axis}
\end{tikzpicture}
\end{subfigure}
% Second subfigure
\begin{subfigure}[b]{\textwidth}
\centering
\begin{tikzpicture}[scale=0.9] 
\begin{axis}[
ybar,
ylabel={F1 top-10},
ymin=0,
ymax=0.25,
width=\textwidth,
bar width=2ex,
height=4cm,  % Adjust the height to fit the vertical arrangement
xtick=data,
xticklabels={Small, Medium, Large , Real-first , Real },
xtick pos=bottom,
ytick pos=left,
enlarge x limits=0.1,
yticklabel style={/pgf/number format/fixed}
]
\addplot [fill=blue]   coordinates {(0,0.10) (1,0.07) (2,0.06) (3,0.05) (4,0.05)};
\addplot [fill=red]    coordinates {(0,0.08) (1,0.03) (2,0.01) (3,0.01) (4,0.01)};
\addplot [fill=green]  coordinates {(0,0.22) (1,0.12) (2,0.04) (3,0.04) (4,0.02)};
\addplot [fill=yellow] coordinates {(0,0.09) (1,0.06) (2,0.04) (3,0.05) (4,0.01)};
\addplot [fill=orange] coordinates {(0,0.11) (1,0.07) (2,0.05) (3,0.06) (4,0.04)};
\end{axis}
\end{tikzpicture}
\end{subfigure}
\begin{subfigure}[b]{\textwidth}
\centering
\begin{tikzpicture}[scale=0.9] 
\begin{axis}[
ybar,
ylabel={F1 dynamic},
ymin=0,
ymax=0.27,
width=\textwidth,
bar width=2ex,
height=4cm,  % Adjust the height to fit the vertical arrangement
xtick=data,
xticklabels={Small, Medium, Large , Real-first , Real },
xtick pos=bottom,
ytick pos=left,
enlarge x limits=0.1,
yticklabel style={/pgf/number format/fixed}
]
\addplot [fill=blue]   coordinates {(0,0.16) (1,0.12) (2,0.09) (3,0.09) (4,0.09)};
\addplot [fill=red]    coordinates {(0,0.16) (1,0.10) (2,0.07) (3,0.06) (4,0.06)};
\addplot [fill=green]  coordinates {(0,0.25) (1,0.15) (2,0.07) (3,0.08) (4,0.04)};
\addplot [fill=yellow] coordinates {(0,0.16) (1,0.12) (2,0.08) (3,0.09) (4,0.06)};
\addplot [fill=orange] coordinates {(0,0.18) (1,0.13) (2,0.09) (3,0.10) (4,0.07)};
\end{axis}
\end{tikzpicture}
\end{subfigure}
\caption{F1 for frequency statistics on \firstdataset.}
\label{fig:f1-wikies}
\vspace{1em}
\end{figure}

\begin{figure}[H]
\centering
% First subfigure
\begin{subfigure}[b]{\textwidth}
\centering
\begin{tikzpicture}[scale=0.9] 
\begin{axis}[
hide axis,
xmin=0,
xmax=10,
ymin=0,
ymax=10,
legend columns=5,
legend style={at={(0.5,1)},anchor=north,draw=none,fill=none,legend cell align=left, xshift=2cm},
legend image post style={scale=1},  % Add this line to scale down the legend images
legend entries={Random, Entity Freq., Inverse Entity Freq., Relation Freq., Inverse Relation Freq.}
]
\addlegendimage{fill=blue,area legend}
\addlegendimage{fill=red,area legend}
\addlegendimage{fill=green,area legend}
\addlegendimage{fill=yellow,area legend}
\addlegendimage{fill=orange,area legend}
\end{axis}
\end{tikzpicture}
\begin{tikzpicture}[scale=0.9] 
\begin{axis}[
ybar,
ylabel={MAP-5},
ymin=0,
ymax=0.12,
width=\textwidth,
bar width=2ex,
height=4cm,  % Adjust the height to fit the vertical arrangement
xtick=data,
xticklabels={Small, Medium, Large , Real-first , Real },
xtick pos=bottom,
ytick pos=left,
enlarge x limits=0.1,
yticklabel style={/pgf/number format/fixed}
]
\addplot [fill=blue]   coordinates {(0,0.02) (1,0.01) (2,0.01) (3,0.01) (4,0.01)};
\addplot [fill=red]    coordinates {(0,0.01) (1,0.01) (2,0.01) (3,0.01) (4,0.01)};
\addplot [fill=green]  coordinates {(0,0.05) (1,0.03) (2,0.01) (3,0.01) (4,0.01)};
\addplot [fill=yellow] coordinates {(0,0.02) (1,0.01) (2,0.01) (3,0.01) (4,0.01)};
\addplot [fill=orange] coordinates {(0,0.02) (1,0.01) (2,0.01) (3,0.01) (4,0.01)};
\end{axis}
\end{tikzpicture}
\end{subfigure}
% Second subfigure
\begin{subfigure}[b]{\textwidth}
\centering
\begin{tikzpicture}[scale=0.9] 
\begin{axis}[
ybar,
ylabel={MAP-10},
ymin=0,
ymax=0.12,
width=\textwidth,
bar width=2ex,
height=4cm,  % Adjust the height to fit the vertical arrangement
xtick=data,
xticklabels={Small, Medium, Large , Real-first , Real },
xtick pos=bottom,
ytick pos=left,
enlarge x limits=0.1,
yticklabel style={/pgf/number format/fixed}
]
\addplot [fill=blue]   coordinates {(0,0.03) (1,0.02) (2,0.01) (3,0.01) (4,0.01)};
\addplot [fill=red]    coordinates {(0,0.01) (1,0.01) (2,0.01) (3,0.01) (4,0.01)};
\addplot [fill=green]  coordinates {(0,0.08) (1,0.05) (2,0.01) (3,0.01) (4,0.01)};
\addplot [fill=yellow] coordinates {(0,0.03) (1,0.02) (2,0.01) (3,0.01) (4,0.01)};
\addplot [fill=orange] coordinates {(0,0.04) (1,0.02) (2,0.01) (3,0.02) (4,0.01)};
\end{axis}
\end{tikzpicture}
\end{subfigure}
\begin{subfigure}[b]{\textwidth}
\centering
\begin{tikzpicture}[scale=0.9] 
\begin{axis}[
ybar,
ylabel={MAP dynamic},
ymin=0,
ymax=0.15,
width=\textwidth,
bar width=2ex,
height=4cm,  % Adjust the height to fit the vertical arrangement
xtick=data,
xticklabels={Small, Medium, Large , Real-first , Real },
xtick pos=bottom,
ytick pos=left,
enlarge x limits=0.1,
yticklabel style={/pgf/number format/fixed}
]
\addplot [fill=blue]   coordinates {(0,0.05) (1,0.03) (2,0.02) (3,0.02) (4,0.02)};
\addplot [fill=red]    coordinates {(0,0.04) (1,0.02) (2,0.01) (3,0.01) (4,0.01)};
\addplot [fill=green]  coordinates {(0,0.10) (1,0.06) (2,0.02) (3,0.01) (4,0.01)};
\addplot [fill=yellow] coordinates {(0,0.06) (1,0.04) (2,0.02) (3,0.02) (4,0.01)};
\addplot [fill=orange] coordinates {(0,0.06) (1,0.04) (2,0.03) (3,0.03) (4,0.01)};
\end{axis}
\end{tikzpicture}
\end{subfigure}
\caption{MAP for frequency statistics on \firstdataset.}
\label{fig:map-wikies}
\vspace{1em}
\end{figure}
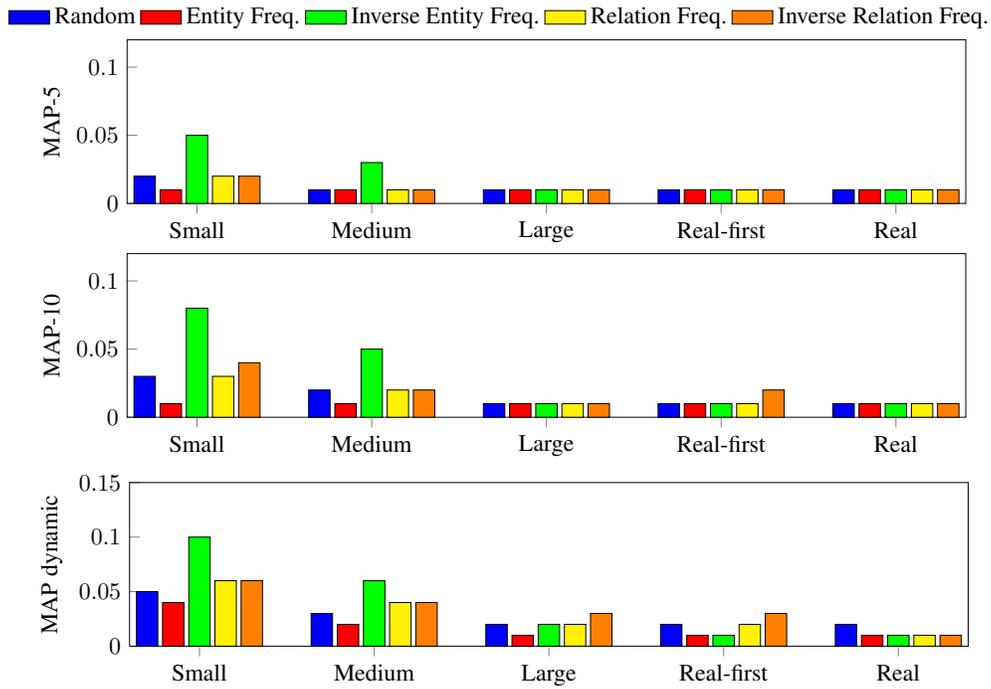
\begin{figure}[H]
\centering
% First subfigure
\begin{subfigure}[b]{\textwidth}
\centering
\begin{tikzpicture}[scale=0.9] 
\begin{axis}[
hide axis,
xmin=0,
xmax=10,
ymin=0,
ymax=10,
legend columns=5,
legend style={at={(0.5,1)},anchor=north,draw=none,fill=none,legend cell align=left, xshift=2cm},
legend image post style={scale=1},  % Add this line to scale down the legend images
legend entries={Random, Entity Freq., Inverse Entity Freq., Relation Freq., Inverse Relation Freq.}
]
\addlegendimage{fill=blue,area legend}
\addlegendimage{fill=red,area legend}
\addlegendimage{fill=green,area legend}
\addlegendimage{fill=yellow,area legend}
\addlegendimage{fill=orange,area legend}
\end{axis}
\end{tikzpicture}
\begin{tikzpicture}[scale=0.9] 
\begin{axis}[
ybar,
ylabel={F1 top-5},
ymin=0,
ymax=0.20,
width=\textwidth,
bar width=2ex,
height=4cm,  % Adjust the height to fit the vertical arrangement
xtick=data,
xticklabels={Small, Medium, Large , Real-first , Real },
xtick pos=bottom,
ytick pos=left,
enlarge x limits=0.1,
yticklabel style={/pgf/number format/fixed}
]
\addplot [fill=blue]   coordinates {(0,0.07) (1,0.06) (2,0.05) (3,0.05) (4,0.09)};
\addplot [fill=red]    coordinates {(0,0.01) (1,0.01) (2,0.01) (3,0.01) (4,0.01)};
\addplot [fill=green]  coordinates {(0,0.14) (1,0.08) (2,0.03) (3,0.02) (4,0.03)};
\addplot [fill=yellow] coordinates {(0,0.08) (1,0.07) (2,0.06) (3,0.06) (4,0.01)};
\addplot [fill=orange] coordinates {(0,0.05) (1,0.04) (2,0.04) (3,0.04) (4,0.05)};
\end{axis}
\end{tikzpicture}
\end{subfigure}
% Second subfigure
\begin{subfigure}[b]{\textwidth}
\centering
\begin{tikzpicture}[scale=0.9] 
\begin{axis}[
ybar,
ylabel={F1 top-10},
ymin=0,
ymax=0.20,
width=\textwidth,
bar width=2ex,
height=4cm,  % Adjust the height to fit the vertical arrangement
xtick=data,
xticklabels={Small, Medium, Large , Real-first , Real },
xtick pos=bottom,
ytick pos=left,
enlarge x limits=0.1,
yticklabel style={/pgf/number format/fixed}
]
\addplot [fill=blue]   coordinates {(0,0.13) (1,0.10) (2,0.09) (3,0.09) (4,0.09)};
\addplot [fill=red]    coordinates {(0,0.04) (1,0.01) (2,0.01) (3,0.01) (4,0.01)};
\addplot [fill=green]  coordinates {(0,0.18) (1,0.11) (2,0.05) (3,0.03) (4,0.03)};
\addplot [fill=yellow] coordinates {(0,0.15) (1,0.12) (2,0.11) (3,0.11) (4,0.01)};
\addplot [fill=orange] coordinates {(0,0.08) (1,0.07) (2,0.06) (3,0.06) (4,0.05)};
\end{axis}
\end{tikzpicture}
\end{subfigure}
\begin{subfigure}[b]{\textwidth}
\centering
\begin{tikzpicture}[scale=0.9] 
\begin{axis}[
ybar,
ylabel={F1 dynamic},
ymin=0,
ymax=0.27,
width=\textwidth,
bar width=2ex,
height=4cm,  % Adjust the height to fit the vertical arrangement
xtick=data,
xticklabels={Small, Medium, Large , Real-first , Real },
xtick pos=bottom,
ytick pos=left,
enlarge x limits=0.1,
yticklabel style={/pgf/number format/fixed}
]
\addplot [fill=blue]   coordinates {(0,0.20) (1,0.17) (2,0.15) (3,0.14) (4,0.14)};
\addplot [fill=red]    coordinates {(0,0.15) (1,0.10) (2,0.08) (3,0.07) (4,0.07)};
\addplot [fill=green]  coordinates {(0,0.23) (1,0.15) (2,0.09) (3,0.06) (4,0.06)};
\addplot [fill=yellow] coordinates {(0,0.25) (1,0.21) (2,0.19) (3,0.19) (4,0.08)};
\addplot [fill=orange] coordinates {(0,0.17) (1,0.14) (2,0.12) (3,0.12) (4,0.11)};
\end{axis}
\end{tikzpicture}
\end{subfigure}
\caption{F1 for frequency statistics on \seconddataset.}
\label{fig:f1-wikies}
\vspace{1em}
\end{figure}
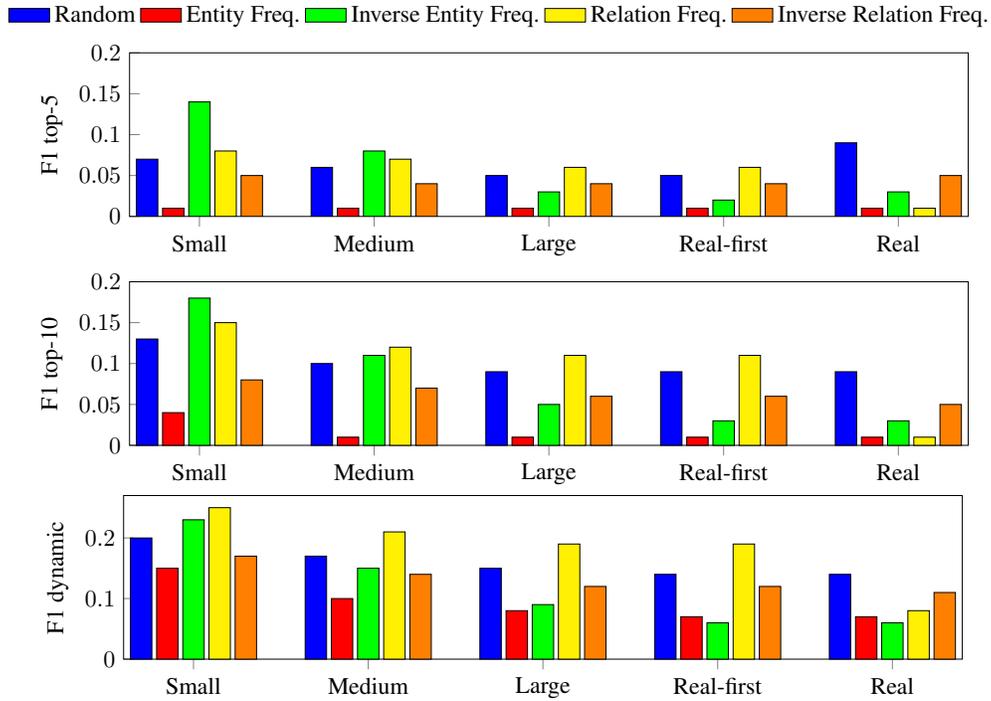

\begin{figure}[H]
\centering
% First subfigure
\begin{subfigure}[b]{\textwidth}
\centering
\begin{tikzpicture}[scale=0.9] 
\begin{axis}[
hide axis,
xmin=0,
xmax=10,
ymin=0,
ymax=10,
legend columns=5,
legend style={at={(0.5,1)},anchor=north,draw=none,fill=none,legend cell align=left, xshift=2cm},
legend image post style={scale=1},  % Add this line to scale down the legend images
legend entries={Random, Entity Freq., Inverse Entity Freq., Relation Freq., Inverse Relation Freq.}
]
\addlegendimage{fill=blue,area legend}
\addlegendimage{fill=red,area legend}
\addlegendimage{fill=green,area legend}
\addlegendimage{fill=yellow,area legend}
\addlegendimage{fill=orange,area legend}
\end{axis}
\end{tikzpicture}
\begin{tikzpicture}[scale=0.9] 
\begin{axis}[
ybar,
ylabel={MAP-5},
ymin=0,
ymax=0.12,
width=\textwidth,
bar width=2ex,
height=4cm,  % Adjust the height to fit the vertical arrangement
xtick=data,
xticklabels={Small, Medium, Large , Real-first , Real },
xtick pos=bottom,
ytick pos=left,
enlarge x limits=0.1,
yticklabel style={/pgf/number format/fixed}
]
\addplot [fill=blue]   coordinates {(0,0.02) (1,0.02) (2,0.01) (3,0.01) (4,0.01)};
\addplot [fill=red]    coordinates {(0,0.01) (1,0.01) (2,0.01) (3,0.01) (4,0.01)};
\addplot [fill=green]  coordinates {(0,0.07) (1,0.03) (2,0.01) (3,0.01) (4,0.01)};
\addplot [fill=yellow] coordinates {(0,0.03) (1,0.02) (2,0.02) (3,0.02) (4,0.01)};
\addplot [fill=orange] coordinates {(0,0.01) (1,0.01) (2,0.01) (3,0.01) (4,0.01)};
\end{axis}
\end{tikzpicture}
\end{subfigure}
% Second subfigure
\begin{subfigure}[b]{\textwidth}
\centering
\begin{tikzpicture}[scale=0.9] 
\begin{axis}[
ybar,
ylabel={MAP-10},
ymin=0,
ymax=0.12,
width=\textwidth,
bar width=2ex,
height=4cm,  % Adjust the height to fit the vertical arrangement
xtick=data,
xticklabels={Small, Medium, Large , Real-first , Real },
xtick pos=bottom,
ytick pos=left,
enlarge x limits=0.1,
yticklabel style={/pgf/number format/fixed}
]
\addplot [fill=blue]   coordinates {(0,0.04) (1,0.03) (2,0.02) (3,0.02) (4,0.01)};
\addplot [fill=red]    coordinates {(0,0.01) (1,0.01) (2,0.01) (3,0.01) (4,0.01)};
\addplot [fill=green]  coordinates {(0,0.09) (1,0.04) (2,0.01) (3,0.01) (4,0.01)};
\addplot [fill=yellow] coordinates {(0,0.05) (1,0.04) (2,0.03) (3,0.04) (4,0.01)};
\addplot [fill=orange] coordinates {(0,0.02) (1,0.02) (2,0.01) (3,0.02) (4,0.01)};
\end{axis}
\end{tikzpicture}
\end{subfigure}
\begin{subfigure}[b]{\textwidth}
\centering
\begin{tikzpicture}[scale=0.9] 
\begin{axis}[
ybar,
ylabel={MAP dynamic},
ymin=0,
ymax=0.15,
width=\textwidth,
bar width=2ex,
height=4cm,  % Adjust the height to fit the vertical arrangement
xtick=data,
xticklabels={Small, Medium, Large , Real-first , Real },
xtick pos=bottom,
ytick pos=left,
enlarge x limits=0.1,
yticklabel style={/pgf/number format/fixed}
]
\addplot [fill=blue]   coordinates {(0,0.07) (1,0.05) (2,0.04) (3,0.04) (4,0.04)};
\addplot [fill=red]    coordinates {(0,0.02) (1,0.01) (2,0.01) (3,0.01) (4,0.01)};
\addplot [fill=green]  coordinates {(0,0.12) (1,0.06) (2,0.02) (3,0.01) (4,0.01)};
\addplot [fill=yellow] coordinates {(0,0.10) (1,0.08) (2,0.07) (3,0.07) (4,0.01)};
\addplot [fill=orange] coordinates {(0,0.05) (1,0.04) (2,0.03) (3,0.03) (4,0.02)};
\end{axis}
\end{tikzpicture}
\end{subfigure}
\caption{MAP for frequency statistics on \seconddataset.}
\label{fig:map-wikies}
\vspace{1em}
\end{figure}
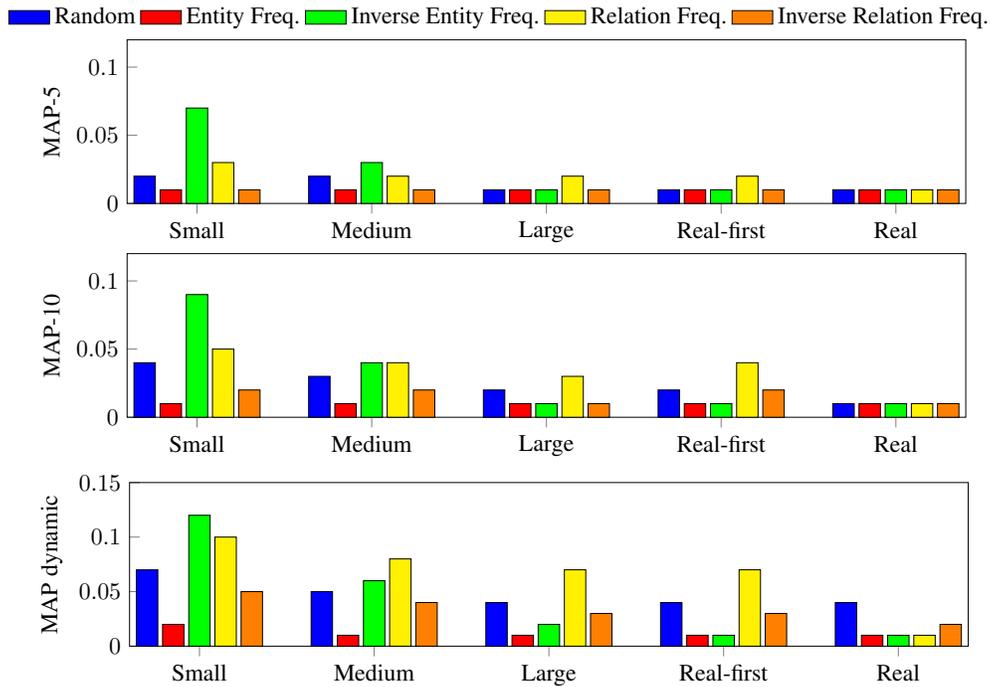
\begin{figure}[H]
\centering
% First subfigure
\begin{subfigure}[b]{\textwidth}
\centering
\begin{tikzpicture}[scale=0.9] 
\begin{axis}[
hide axis,
xmin=0,
xmax=10,
ymin=0,
ymax=10,
legend columns=5,
legend style={at={(0.5,1)},anchor=north,draw=none,fill=none,legend cell align=left, xshift=2cm},
legend image post style={scale=1},  % Add this line to scale down the legend images
legend entries={Random, Entity Freq., Inverse Entity Freq., Relation Freq., Inverse Relation Freq.}
]
\addlegendimage{fill=blue,area legend}
\addlegendimage{fill=red,area legend}
\addlegendimage{fill=green,area legend}
\addlegendimage{fill=yellow,area legend}
\addlegendimage{fill=orange,area legend}
\end{axis}
\end{tikzpicture}
\begin{tikzpicture}[scale=0.9] 
\begin{axis}[
ybar,
ylabel={F1 top-5},
ymin=0,
ymax=0.22,
width=\textwidth,
bar width=2ex,
height=4cm,  % Adjust the height to fit the vertical arrangement
xtick=data,
xticklabels={Small, Medium, Large , Real-first , Real },
xtick pos=bottom,
ytick pos=left,
enlarge x limits=0.1,
yticklabel style={/pgf/number format/fixed}
]
\addplot [fill=blue]   coordinates {(0,0.08) (1,0.06) (2,0.05) (3,0.06) (4,0.06)};
\addplot [fill=red]    coordinates {(0,0.05) (1,0.04) (2,0.03) (3,0.02) (4,0.02)};
\addplot [fill=green]  coordinates {(0,0.13) (1,0.08) (2,0.04) (3,0.03) (4,0.02)};
\addplot [fill=yellow] coordinates {(0,0.03) (1,0.02) (2,0.02) (3,0.02) (4,0.01)};
\addplot [fill=orange] coordinates {(0,0.08) (1,0.07) (2,0.06) (3,0.07) (4,0.05)};
\end{axis}
\end{tikzpicture}
\end{subfigure}
% Second subfigure
\begin{subfigure}[b]{\textwidth}
\centering
\begin{tikzpicture}[scale=0.9] 
\begin{axis}[
ybar,
ylabel={F1 top-10},
ymin=0,
ymax=0.22,
width=\textwidth,
bar width=2ex,
height=4cm,  % Adjust the height to fit the vertical arrangement
xtick=data,
xticklabels={Small, Medium, Large , Real-first , Real },
xtick pos=bottom,
ytick pos=left,
enlarge x limits=0.1,
yticklabel style={/pgf/number format/fixed}
]
\addplot [fill=blue]   coordinates {(0,0.12) (1,0.10) (2,0.09) (3,0.09) (4,0.09)};
\addplot [fill=red]    coordinates {(0,0.15) (1,0.11) (2,0.09) (3,0.06) (4,0.06)};
\addplot [fill=green]  coordinates {(0,0.19) (1,0.12) (2,0.06) (3,0.04) (4,0.04)};
\addplot [fill=yellow] coordinates {(0,0.06) (1,0.05) (2,0.04) (3,0.04) (4,0.06)};
\addplot [fill=orange] coordinates {(0,0.14) (1,0.12) (2,0.11) (3,0.13) (4,0.09)};
\end{axis}
\end{tikzpicture}
\end{subfigure}
\begin{subfigure}[b]{\textwidth}
\centering
\begin{tikzpicture}[scale=0.9] 
\begin{axis}[
ybar,
ylabel={F1 dynamic},
ymin=0,
ymax=0.27,
width=\textwidth,
bar width=2ex,
height=4cm,  % Adjust the height to fit the vertical arrangement
xtick=data,
xticklabels={Small, Medium, Large , Real-first , Real },
xtick pos=bottom,
ytick pos=left,
enlarge x limits=0.1,
yticklabel style={/pgf/number format/fixed}
]
\addplot [fill=blue]   coordinates {(0,0.18) (1,0.14) (2,0.13) (3,0.13) (4,0.13)};
\addplot [fill=red]    coordinates {(0,0.25) (1,0.21) (2,0.18) (3,0.16) (4,0.16)};
\addplot [fill=green]  coordinates {(0,0.22) (1,0.15) (2,0.09) (3,0.06) (4,0.05)};
\addplot [fill=yellow] coordinates {(0,0.10) (1,0.08) (2,0.06) (3,0.06) (4,0.16)};
\addplot [fill=orange] coordinates {(0,0.22) (1,0.18) (2,0.16) (3,0.18) (4,0.12)};
\end{axis}
\end{tikzpicture}
\end{subfigure}
\caption{F1 for frequency statistics on \thirddataset.}
\label{fig:f1-wikies}
\vspace{1em}
\end{figure}
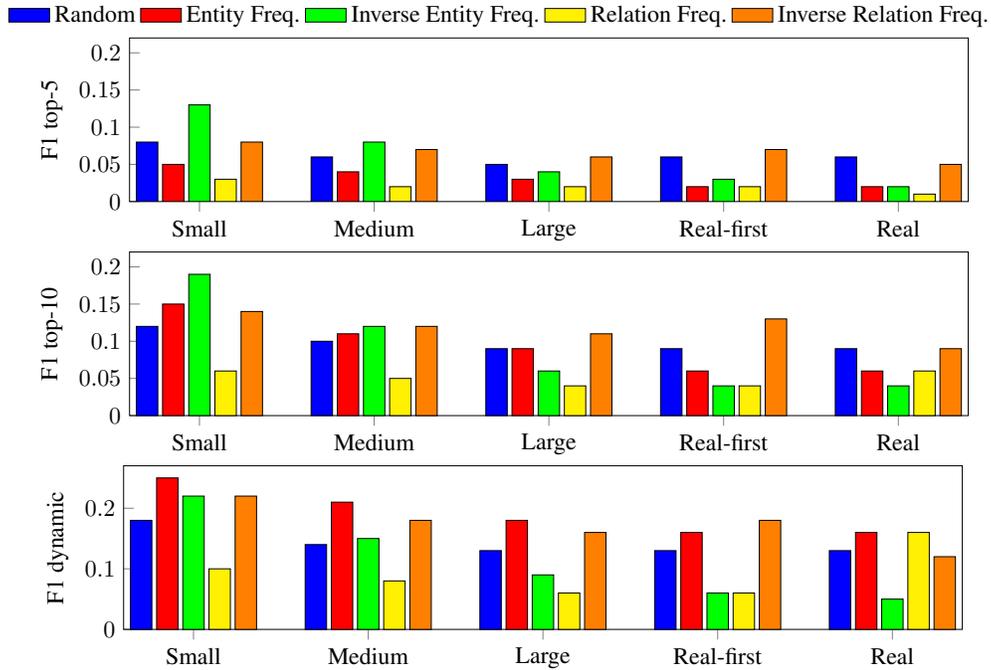

\begin{figure}[H]
\centering
% First subfigure
\begin{subfigure}[b]{\textwidth}
\centering
\begin{tikzpicture}[scale=0.9] 
\begin{axis}[
hide axis,
xmin=0,
xmax=10,
ymin=0,
ymax=10,
legend columns=5,
legend style={at={(0.5,1)},anchor=north,draw=none,fill=none,legend cell align=left, xshift=2cm},
legend image post style={scale=1},  % Add this line to scale down the legend images
legend entries={Random, Entity Freq., Inverse Entity Freq., Relation Freq., Inverse Relation Freq.}
]
\addlegendimage{fill=blue,area legend}
\addlegendimage{fill=red,area legend}
\addlegendimage{fill=green,area legend}
\addlegendimage{fill=yellow,area legend}
\addlegendimage{fill=orange,area legend}
\end{axis}
\end{tikzpicture}
\begin{tikzpicture}[scale=0.9] 
\begin{axis}[
ybar,
ylabel={MAP-5},
ymin=0,
ymax=0.12,
width=\textwidth,
bar width=2ex,
height=4cm,  % Adjust the height to fit the vertical arrangement
xtick=data,
xticklabels={Small, Medium, Large , Real-first , Real },
xtick pos=bottom,
ytick pos=left,
enlarge x limits=0.1,
yticklabel style={/pgf/number format/fixed}
]
\addplot [fill=blue]   coordinates {(0,0.03) (1,0.02) (2,0.02) (3,0.02) (4,0.02)};
\addplot [fill=red]    coordinates {(0,0.01) (1,0.01) (2,0.01) (3,0.01) (4,0.01)};
\addplot [fill=green]  coordinates {(0,0.06) (1,0.03) (2,0.01) (3,0.01) (4,0.01)};
\addplot [fill=yellow] coordinates {(0,0.01) (1,0.01) (2,0.01) (3,0.01) (4,0.01)};
\addplot [fill=orange] coordinates {(0,0.03) (1,0.02) (2,0.02) (3,0.02) (4,0.02)};
\end{axis}
\end{tikzpicture}
\end{subfigure}
% Second subfigure
\begin{subfigure}[b]{\textwidth}
\centering
\begin{tikzpicture}[scale=0.9] 
\begin{axis}[
ybar,
ylabel={MAP-10},
ymin=0,
ymax=0.12,
width=\textwidth,
bar width=2ex,
height=4cm,  % Adjust the height to fit the vertical arrangement
xtick=data,
xticklabels={Small, Medium, Large , Real-first , Real },
xtick pos=bottom,
ytick pos=left,
enlarge x limits=0.1,
yticklabel style={/pgf/number format/fixed}
]
\addplot [fill=blue]   coordinates {(0,0.04) (1,0.03) (2,0.03) (3,0.03) (4,0.03)};
\addplot [fill=red]    coordinates {(0,0.04) (1,0.03) (2,0.02) (3,0.01) (4,0.01)};
\addplot [fill=green]  coordinates {(0,0.08) (1,0.05) (2,0.02) (3,0.01) (4,0.01)};
\addplot [fill=yellow] coordinates {(0,0.02) (1,0.01) (2,0.01) (3,0.01) (4,0.01)};
\addplot [fill=orange] coordinates {(0,0.05) (1,0.04) (2,0.04) (3,0.04) (4,0.03)};
\end{axis}
\end{tikzpicture}
\end{subfigure}
\begin{subfigure}[b]{\textwidth}
\centering
\begin{tikzpicture}[scale=0.9] 
\begin{axis}[
ybar,
ylabel={MAP dynamic},
ymin=0,
ymax=0.15,
width=\textwidth,
bar width=2ex,
height=4cm,  % Adjust the height to fit the vertical arrangement
xtick=data,
xticklabels={Small, Medium, Large , Real-first , Real },
xtick pos=bottom,
ytick pos=left,
enlarge x limits=0.1,
yticklabel style={/pgf/number format/fixed}
]
\addplot [fill=blue]   coordinates {(0,0.07) (1,0.05) (2,0.04) (3,0.04) (4,0.04)};
\addplot [fill=red]    coordinates {(0,0.09) (1,0.06) (2,0.05) (3,0.04) (4,0.04)};
\addplot [fill=green]  coordinates {(0,0.12) (1,0.06) (2,0.03) (3,0.01) (4,0.01)};
\addplot [fill=yellow] coordinates {(0,0.03) (1,0.02) (2,0.02) (3,0.02) (4,0.04)};
\addplot [fill=orange] coordinates {(0,0.08) (1,0.06) (2,0.06) (3,0.06) (4,0.04)};
\end{axis}
\end{tikzpicture}
\end{subfigure}
\caption{MAP for frequency statistics on \thirddataset.}
\label{fig:map-wikies}
\vspace{1em}
\end{figure}
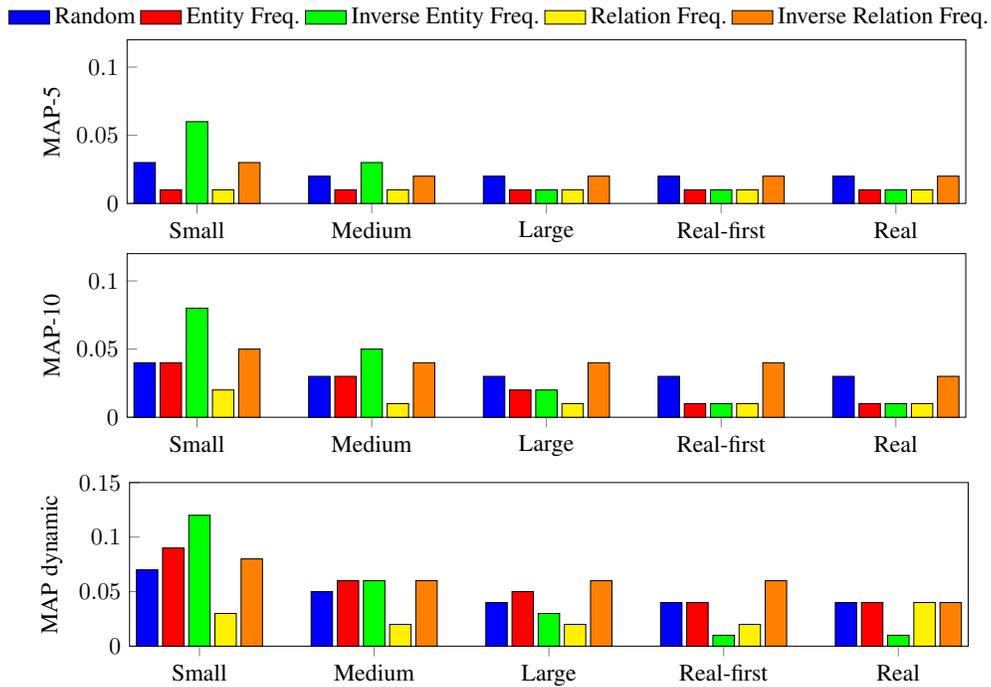
\begin{figure}[H]
\centering
% First subfigure
\begin{subfigure}[b]{\textwidth}
\centering
\begin{tikzpicture}[scale=0.9] 
\begin{axis}[
hide axis,
xmin=0,
xmax=10,
ymin=0,
ymax=10,
legend columns=5,
legend style={at={(0.5,1)},anchor=north,draw=none,fill=none,legend cell align=left, xshift=2cm},
legend image post style={scale=1},  % Add this line to scale down the legend images
legend entries={Random, Entity Freq., Inverse Entity Freq., Relation Freq., Inverse Relation Freq.}
]
\addlegendimage{fill=blue,area legend}
\addlegendimage{fill=red,area legend}
\addlegendimage{fill=green,area legend}
\addlegendimage{fill=yellow,area legend}
\addlegendimage{fill=orange,area legend}
\end{axis}
\end{tikzpicture}
\begin{tikzpicture}[scale=0.9] 
\begin{axis}[
ybar,
ylabel={F1 top-5},
ymin=0,
ymax=0.22,
width=\textwidth,
bar width=2ex,
height=4cm,  % Adjust the height to fit the vertical arrangement
xtick=data,
xticklabels={Small, Medium, Large , Real-first , Real },
xtick pos=bottom,
ytick pos=left,
enlarge x limits=0.1,
yticklabel style={/pgf/number format/fixed}
]
\addplot [fill=blue]   coordinates {(0,0.10) (1,0.08) (2,0.09) (3,0.09) (4,0.09)};
\addplot [fill=red]    coordinates {(0,0.02) (1,0.02) (2,0.02) (3,0.02) (4,0.02)};
\addplot [fill=green]  coordinates {(0,0.15) (1,0.09) (2,0.05) (3,0.10) (4,0.05)};
\addplot [fill=yellow] coordinates {(0,0.06) (1,0.06) (2,0.05) (3,0.08) (4,0.03)};
\addplot [fill=orange] coordinates {(0,0.09) (1,0.09) (2,0.08) (3,0.09) (4,0.08)};
\end{axis}
\end{tikzpicture}
\end{subfigure}
% Second subfigure
\begin{subfigure}[b]{\textwidth}
\centering
\begin{tikzpicture}[scale=0.9] 
\begin{axis}[
ybar,
ylabel={F1 top-10},
ymin=0,
ymax=0.22,
width=\textwidth,
bar width=2ex,
height=4cm,  % Adjust the height to fit the vertical arrangement
xtick=data,
xticklabels={Small, Medium, Large , Real-first , Real },
xtick pos=bottom,
ytick pos=left,
enlarge x limits=0.1,
yticklabel style={/pgf/number format/fixed}
]
\addplot [fill=blue]   coordinates {(0,0.16) (1,0.14) (2,0.14) (3,0.14) (4,0.14)};
\addplot [fill=red]    coordinates {(0,0.09) (1,0.07) (2,0.06) (3,0.06) (4,0.07)};
\addplot [fill=green]  coordinates {(0,0.21) (1,0.14) (2,0.10) (3,0.14) (4,0.09)};
\addplot [fill=yellow] coordinates {(0,0.13) (1,0.12) (2,0.11) (3,0.12) (4,0.06)};
\addplot [fill=orange] coordinates {(0,0.15) (1,0.14) (2,0.14) (3,0.14) (4,0.14)};
\end{axis}
\end{tikzpicture}
\end{subfigure}
\begin{subfigure}[b]{\textwidth}
\centering
\begin{tikzpicture}[scale=0.9] 
\begin{axis}[
ybar,
ylabel={F1 dynamic},
ymin=0,
ymax=0.25,
width=\textwidth,
bar width=2ex,
height=4cm,  % Adjust the height to fit the vertical arrangement
xtick=data,
xticklabels={Small, Medium, Large , Real-first , Real },
xtick pos=bottom,
ytick pos=left,
enlarge x limits=0.1,
yticklabel style={/pgf/number format/fixed}
]
\addplot [fill=blue]   coordinates {(0,0.19) (1,0.18) (2,0.18) (3,0.17) (4,0.17)};
\addplot [fill=red]    coordinates {(0,0.13) (1,0.11) (2,0.10) (3,0.09) (4,0.10)};
\addplot [fill=green]  coordinates {(0,0.24) (1,0.17) (2,0.13) (3,0.20) (4,0.12)};
\addplot [fill=yellow] coordinates {(0,0.18) (1,0.16) (2,0.15) (3,0.17) (4,0.10)};
\addplot [fill=orange] coordinates {(0,0.19) (1,0.18) (2,0.18) (3,0.18) (4,0.17)};
\end{axis}
\end{tikzpicture}
\end{subfigure}
\caption{F1 for frequency statistics on \fourthdataset.}
\label{fig:f1-wikies}
\vspace{1em}
\end{figure}
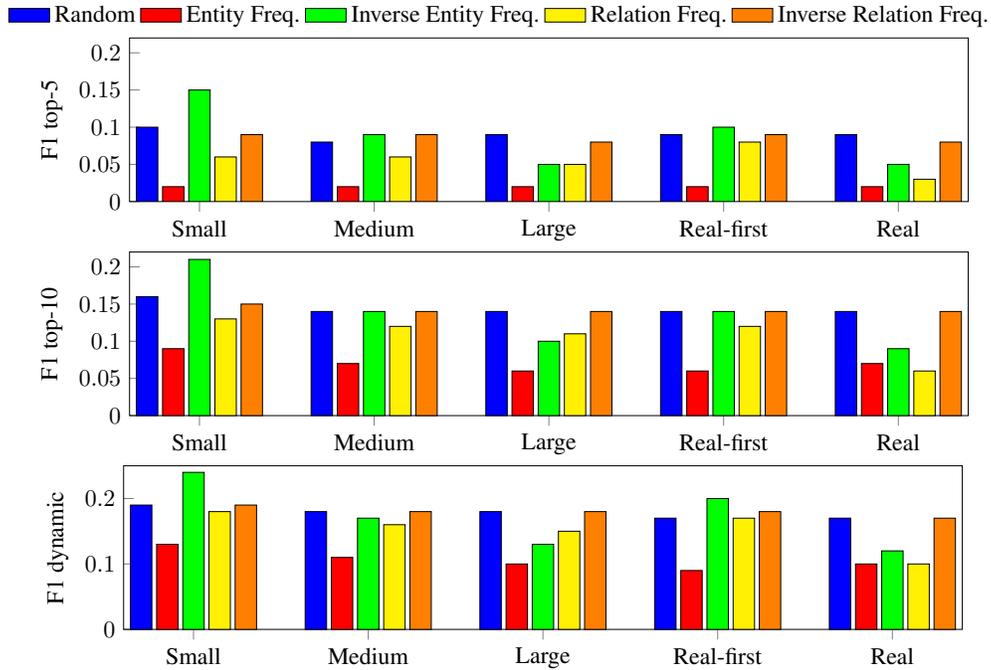

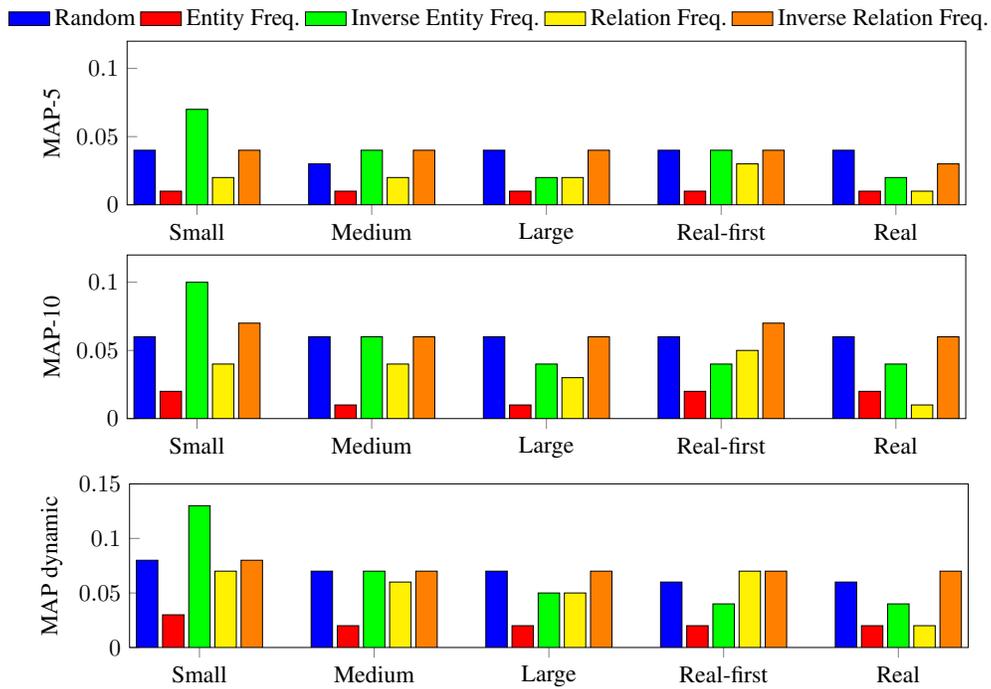
\begin{figure}[H]
\centering
% First subfigure
\begin{subfigure}[b]{\textwidth}
\centering
\begin{tikzpicture}[scale=0.9] 
\begin{axis}[
hide axis,
xmin=0,
xmax=10,
ymin=0,
ymax=10,
legend columns=5,
legend style={at={(0.5,1)},anchor=north,draw=none,fill=none,legend cell align=left, xshift=2cm},
legend image post style={scale=1},  % Add this line to scale down the legend images
legend entries={Random, Entity Freq., Inverse Entity Freq., Relation Freq., Inverse Relation Freq.}
]
\addlegendimage{fill=blue,area legend}
\addlegendimage{fill=red,area legend}
\addlegendimage{fill=green,area legend}
\addlegendimage{fill=yellow,area legend}
\addlegendimage{fill=orange,area legend}
\end{axis}
\end{tikzpicture}
\begin{tikzpicture}[scale=0.9] 
\begin{axis}[
ybar,
ylabel={MAP-5},
ymin=0,
ymax=0.12,
width=\textwidth,
bar width=2ex,
height=4cm,  % Adjust the height to fit the vertical arrangement
xtick=data,
xticklabels={Small, Medium, Large , Real-first , Real },
xtick pos=bottom,
ytick pos=left,
enlarge x limits=0.1,
yticklabel style={/pgf/number format/fixed}
]
\addplot [fill=blue]   coordinates {(0,0.04) (1,0.03) (2,0.04) (3,0.04) (4,0.04)};
\addplot [fill=red]    coordinates {(0,0.01) (1,0.01) (2,0.01) (3,0.01) (4,0.01)};
\addplot [fill=green]  coordinates {(0,0.07) (1,0.04) (2,0.02) (3,0.04) (4,0.02)};
\addplot [fill=yellow] coordinates {(0,0.02) (1,0.02) (2,0.02) (3,0.03) (4,0.01)};
\addplot [fill=orange] coordinates {(0,0.04) (1,0.04) (2,0.04) (3,0.04) (4,0.03)};
\end{axis}
\end{tikzpicture}
\end{subfigure}
% Second subfigure
\begin{subfigure}[b]{\textwidth}
\centering
\begin{tikzpicture}[scale=0.9] 
\begin{axis}[
ybar,
ylabel={MAP-10},
ymin=0,
ymax=0.12,
width=\textwidth,
bar width=2ex,
height=4cm,  % Adjust the height to fit the vertical arrangement
xtick=data,
xticklabels={Small, Medium, Large , Real-first , Real },
xtick pos=bottom,
ytick pos=left,
enlarge x limits=0.1,
yticklabel style={/pgf/number format/fixed}
]
\addplot [fill=blue]   coordinates {(0,0.06) (1,0.06) (2,0.06) (3,0.06) (4,0.06)};
\addplot [fill=red]    coordinates {(0,0.02) (1,0.01) (2,0.01) (3,0.02) (4,0.02)};
\addplot [fill=green]  coordinates {(0,0.10) (1,0.06) (2,0.04) (3,0.04) (4,0.04)};
\addplot [fill=yellow] coordinates {(0,0.04) (1,0.04) (2,0.03) (3,0.05) (4,0.01)};
\addplot [fill=orange] coordinates {(0,0.07) (1,0.06) (2,0.06) (3,0.07) (4,0.06)};
\end{axis}
\end{tikzpicture}
\end{subfigure}
\begin{subfigure}[b]{\textwidth}
\centering
\begin{tikzpicture}[scale=0.9] 
\begin{axis}[
ybar,
ylabel={MAP dynamic},
ymin=0,
ymax=0.15,
width=\textwidth,
bar width=2ex,
height=4cm,  % Adjust the height to fit the vertical arrangement
xtick=data,
xticklabels={Small, Medium, Large , Real-first , Real },
xtick pos=bottom,
ytick pos=left,
enlarge x limits=0.1,
yticklabel style={/pgf/number format/fixed}
]
\addplot [fill=blue]   coordinates {(0,0.08) (1,0.07) (2,0.07) (3,0.06) (4,0.06)};
\addplot [fill=red]    coordinates {(0,0.03) (1,0.02) (2,0.02) (3,0.02) (4,0.02)};
\addplot [fill=green]  coordinates {(0,0.13) (1,0.07) (2,0.05) (3,0.04) (4,0.04)};
\addplot [fill=yellow] coordinates {(0,0.07) (1,0.06) (2,0.05) (3,0.07) (4,0.02)};
\addplot [fill=orange] coordinates {(0,0.08) (1,0.07) (2,0.07) (3,0.07) (4,0.07)};
\end{axis}
\end{tikzpicture}
\end{subfigure}
\caption{MAP for frequency statistics on \fourthdataset.}
\label{fig:map-wikies}
\vspace{1em}
\end{figure}

\end{document}